\documentclass[10pt,twocolumn,twoside]{IEEEtran}
\usepackage{bbm}
\usepackage{amssymb,graphicx,lineno}
\usepackage{rawfonts,amsfonts,amssymb,psfrag,color}
\usepackage{amsfonts}
\usepackage{amsmath,epsfig}
\usepackage{rawfonts,graphicx,amsfonts,amssymb}
\usepackage{bm}
\usepackage{url}
\usepackage{algorithm, algorithmic}

\begin{document}

\title{Coordinate Descent Algorithms for Phase Retrieval}

\author{Wen-Jun Zeng,~\IEEEmembership{Member,~IEEE}, and H. C. So,~\IEEEmembership{Fellow,~IEEE}
\thanks{The authors are with the Department of Electronic Engineering, City University of Hong Kong, Hong Kong.
(E-mail: wenjzeng@cityu.edu.hk, hcso@ee.cityu.edu.hk).}}

\maketitle

\begin{abstract}
Phase retrieval aims at recovering a complex-valued signal from magnitude-only measurements, which attracts much attention since it has numerous applications in many disciplines. However, phase recovery involves solving a system of quadratic equations, indicating that it is a challenging nonconvex optimization problem. To tackle phase retrieval in an effective and efficient manner, we apply coordinate descent (CD) such that a single unknown is solved at each iteration while all other variables are kept fixed. As a result, only minimization of a univariate quartic polynomial is needed which is easily achieved by finding the closed-form roots of a cubic equation. Three computationally simple algorithms referred to as cyclic, randomized and greedy CDs, based on different updating rules, are devised. It is proved that the three CDs globally converge to a stationary point of the nonconvex problem, and specifically, the randomized CD locally converges to the global minimum and attains exact recovery at a geometric rate with high probability if the sample size is large enough. The cyclic and randomized CDs are also modified via minimization of the $\ell_1$-regularized quartic polynomial for phase retrieval of sparse signals. Furthermore, a novel application of the three CDs, namely, blind equalization in digital communications, is proposed. It is demonstrated that the CD methodology is superior to the state-of-the-art techniques in terms of computational efficiency and/or recovery performance.
\end{abstract}


\section{Introduction}
Phase retrieval refers to the recovery of a complex-valued signal from only intensity or squared-magnitude measurements of its linear transformation \cite{Patent,Candes:MC,SPM}. It has been a very active field of research because of its wide applicability in science and engineering, which include areas of optical imaging \cite{SPM}, crystallography \cite{Xray}, electron microscopy \cite{JMiao}, neutron radiography \cite{Nature}, digital communications \cite{Baykal04}, astronomy \cite{Dainty87} and computational biology \cite{Stefik78}. The first model for phase retrieval investigates the problem of recovering a signal from the squared-magnitude of its Fourier transform. To address various applications, the power spectrum measurement model has been extended to different formulations, including the short-time Fourier transform \cite{Balan,STFT}, coded diffraction patterns \cite{CDP}, and random measurements \cite{Candes:MC}, \cite{WF}--\cite{JuSun}. Nevertheless, in all these models, observations of the signal-of-interest (SOI) are obtained via a linear mapping, and we can only measure the intensity.

Early approach to phase retrieval is based on error reduction, which includes the most representative Gerchberg-Saxton (GS) algorithm \cite{GSA} and its modified version proposed by Fienup \cite{Fienup}, as well as other variants \cite{AltMin}--\cite{Elser}. In essence, the error reduction techniques apply the concept of alternating projection. That is, at each iteration, the current SOI estimate is projected onto one constraint set such that the magnitudes of its linear mapping match the observations, and then the signal is projected onto another constraint set to conform to the \emph{a priori} knowledge about its structure \cite{WF,GSA}. This methodology works well in practice but its convergence is unclear because projection onto nonconvex sets is involved. Recently, the guarantee of convergence to global solution for the GS algorithm is proved under the condition of resampling \cite{AltMin}. The number of measurements required in the resampled GS scheme is on the order of $N\log^3 N$ with $N$ being the signal length. Nevertheless, it will be clear later that this sampling complexity is not optimal compared with other advanced methods.

In fact, phase recovery corresponds to a nonconvex optimization problem. To be specific, it requires solving a \emph{system of quadratic equations}, or equivalently, minimizing a multivariate fourth-order polynomial, which is generally known to be NP-hard \cite{WF,JuSun,NP:hard,TWF}. The convex relaxation based methods, including PhaseLift \cite{Candes:MC,PhaseLift} and PhaseCut \cite{PhaseCut}, relax the original nonconvex problem into a convex program. The PhaseLift converts the quadratic equations into linear ones by lifting the $N$-dimensional signal vector to an $N \times N$  rank-one matrix. Then it approximates the minimum rank problem using trace norm minimization, which is convex and can be solved by semidefinite programming (SDP). The sampling complexity of PhaseLift is $\mathcal{O}(N\log N)$, which is lower than that of the resampled alternating projection method \cite{AltMin} and is nearly optimal \cite{PhaseLift}. On the other hand, the PhaseCut recasts phase retrieval as a quadratically constrained quadratic program (QCQP) which is then approximately solved via semidefinite relaxation \cite{MaxCut,QCQP}. It has similar sampling complexity to PhaseLift and both exhibit good retrieval performance. However, the computational load of the SDP based methods is very high, especially when the signal length or number of observations is large, since the PhaseLift and PhaseCut involve matrix variables with $\mathcal{O}(N^2)$ and $\mathcal{O}(M^2)$ elements, respectively, where $M$ is the measurement number. As a result, the convex relaxation approach cannot deal with large-scale problems.

To circumvent the high computational requirement, Wirtinger flow (WF) \cite{WF}, which is essentially a gradient descent technique for complex-valued variables, is developed for minimizing the nonconvex quartic polynomial. In general, the gradient method is only guaranteed to converge to a stationary point of a nonconvex objective function.
In other words, it can trap in a saddle point or local minimum. That is to say, convergence to the global solution is not guaranteed for general nonconvex optimization problems using the gradient descent. Surprisingly, when initiated via a spectral method \cite{WF} and the sample size is $\mathcal{O}(N\log N)$, Cand\`{e}s $et$ $al.$ prove that the WF algorithm converges to the global solution at a geometric rate with high probability. The truncated WF \cite{TWF} further enhances the recovery performance by adaptively selecting a portion of measurements at each iteration while the optimal stepsize for convergence rate acceleration has been derived in \cite{WFOS}. Still, the convergence speed of the gradient-based WF approach is not fast.

In many applications, the SOI is sparse or only contains a few nonzero entries in some basis. Recovering a sparse signal from the intensity-only measurements is called \emph{quadratic compressed sensing}. Like classic compressed sensing based on linear measurements \cite{Candes:SPM,Candes:CS}, the sampling complexity
of phase retrieval can be reduced by exploiting sparsity. Several above-mentioned phase recovery schemes
for non-sparse signals have been adapted to handle sparse SOIs. For example, performing hard-thresholding at each iteration of the GS or Fienup scheme yields the
so-called the sparse Fienup algorithm \cite{SparseFienup}. Similarly, applying a
thresholding operation\footnote{The thresholding operator can be soft or hard.}
to the WF method elicits the thresholded WF algorithm \cite{TCai}. Both will yield desirable solution because either soft- or hard-thresholding enforces the signal to be sparse. Furthermore, by borrowing the idea from orthogonal matching pursuit \cite{OMP,LpGreedy}, a greedy algorithm is designed for sparse phase retrieval in \cite{GESPAR}.

In this work, we develop effective and computationally efficient algorithms
with faster convergence rate for minimizing the nonconvex quartic polynomial in phase retrieval.
Our approach is based on coordinate descent (CD), which adopts the strategy of ``one at a time'' \cite{Wright,Nesterov}. That is, CD solves a multivariate minimization problem by successively finding a single unknown at each iteration while keeping the remaining variables fixed. According to different rules for coordinate
selection, our scheme includes three variants, namely, cyclic, randomized, and greedy CDs.
One motivation using CD for phase retrieval is that the exact minimizer of each coordinate is
easily obtained by finding the roots of a univariate cubic equation. It is believed that the proposed methodology provides a new path to solve phase retrieval and related problems.

We summarize the contributions of this paper as follows.
\begin{itemize}
  \item [(i)] An algorithmic framework including cyclic, randomized, and greedy CDs,
  is proposed to solve the quartic polynomial minimization for phase retrieval.
  The CD algorithm is computationally simple and converges much faster than the
  gradient descent methods such as WF and its variants \cite{WF, TWF, WFOS}.
  \item [(ii)] Theoretically, we prove that the CD globally converges to a stationary point
of the nonconvex problem, where the gradient is non-Lipschitz continuous. It is worth pointing out that the proof is nontrivial because the existing convergence analyses of CD assuming convexity and Lipschitz continuity
\cite{Wright,Nesterov} are not applicable to our problem.
  \item [(iii)] It is proved that the randomized CD locally converges to the global minimum
  at a geometric rate with high probability using $\mathcal{O}(N\log N)$ measurements.
  \item [(iv)] The CD algorithms are extended for phase retrieval of
  sparse signals, where the minimization of the $\ell_1$-regularized quartic polynomial
  is solved.
  \item [(v)] Currently, the applications of phase retrieval mainly focus on imaging. Here, we open up a new use of phase retrieval for blind equalization in digital communications, i.e., removing the adverse effect induced by channel propagation.
\end{itemize}

The remainder of this paper is organized as follows.
The phase retrieval problem is formulated and three CD algorithms are introduced in Section \ref{Sec:CD}.
In Section \ref{Sec:Converge}, two types of convergence, namely, global convergence to a
stationary point and local convergence to the global minimum, are theoretically proved.
Section \ref{Sec:L1CD} presents the $\ell_1$-regularized CD algorithms for sparse phase retrieval.
The application to blind equalization is investigated in Section \ref{Sec:BE}.
Simulation results are provided in Section \ref{Sec:Simulation}. Finally, conclusions are drawn
in Section \ref{Sec:Conclusion}.

We use bold capital upper case and lower case letters to represent
matrices and vectors, respectively. The $i$th element of a vector is expressed as $[\cdot]_i$, and similarly,
the $(i,j)$ entry of a matrix is $[\cdot]_{i,j}$. The identity matrix is denoted by $\pmb I$.
The superscripts $(\cdot)^T$, $(\cdot)^*$ and
$(\cdot)^H$ stand for the transpose, complex conjugate and Hermitian
transpose, respectively. The imaginary unit is ${\rm j}=\sqrt{-1}$ while
$\mathbb{E}[\cdot]$ is expectation operator. The ${\rm Re}(\cdot)$ and ${\rm Im}(\cdot)$ denote the real and imaginary
parts of a complex-valued scalar, vector or matrix. The $\ell_2$-, $\ell_1$-, and $\ell_{\infty}$-norms
of a vector are denoted as $\|\cdot\|$, $\|\cdot\|_1$, and $\|\cdot\|_{\infty}$, respectively.
The inner product is represented as $\left\langle ,\right\rangle $ and $|\cdot|$ means
the absolute value of a real number or the modulus of a complex number. Finally, $\mathbb{R}$ and $\mathbb{C}$ denote
the fields of real and complex numbers, respectively.

\section{CD for Phase Retrieval}\label{Sec:CD}

\subsection{Problem Formulation}

We consider the problem of recovering a complex-valued signal $\pmb x\in\mathbb{C}^N$ from
$M$ phaseless observations $b_m\in\mathbb{R}$:
\begin{equation}\label{Model}
      b_m = \left|\pmb a_m^H\pmb x\right|^2 + \nu_m, \quad m = 1,\cdots,M
\end{equation}
where $\pmb a_m\in\mathbb{C}^N$ are known sampling vectors,
and $\nu_m\in\mathbb{R}$ are additive zero-mean noise terms, and and generally $M >N$. Note that in case of magnitude-only measurements $b_m = \left|\pmb a_m^H\pmb x\right| + \nu_m$, we can convert it to $b_m^2$. The measurements are collected into a vector
$\pmb b = [b_1,\cdots,b_M]^T\in\mathbb{R}^M$. Finding a solution of \eqref{Model} in the noiseless case
refers to solving a quadratic system of equations. Apparently, $\pmb x$ can only be recovered up to a global phase $\phi\in [0,2\pi)$ because ${\rm e}^{{\rm j}\phi}\pmb x$ is also a solution. Adopting the least squares (LS) criterion, $\pmb x$ is determined from:
\begin{equation}\label{Obj}
      \min_{\pmb x\in\mathbb{C}^N} f(\pmb x)
      :=\sum_{m=1}^{M}
      \left( \left|\pmb a_m^H\pmb x\right|^2 - b_m\right)^2.
\end{equation}
When the noise is independent and
identically distributed (i.i.d.) and Gaussian, the LS
estimate given by \eqref{Obj} is equivalent to the maximum likelihood solution.
Nevertheless, the optimization problem of \eqref{Obj} is not easy to solve because it is not only nonlinear
but also nonconvex.

\subsection{Outline of CD}

To derive the CD, we first analyze the structure of the objective function in \eqref{Obj}.
The $m$th ($m = 1,\cdots,M$) term in \eqref{Obj} is
\begin{equation}\label{fm:def}
    f_m(\pmb x) = \left( \left|\pmb a_m^H\pmb x\right|^2 - b_m\right)^2
    = \left( \pmb x^H\pmb A_m\pmb x - b_m\right)^2
\end{equation}
where $\pmb A_m=\pmb a_m\pmb a_m^H\in\mathbb{C}^{N \times N}$ is a rank-one Hermitian matrix.
Our first step is to convert the complex-valued problem into a real-valued one. It will be
revealed shortly why we deal with the real-valued variables instead of complex-valued parameters.
Define the expanded real-valued matrix
\begin{equation}\label{Am:bar}
      \bar{\pmb A}_m=\left[ {\begin{array}{*{20}c}
   {\rm Re}(\pmb A_m) & -{\rm Im}(\pmb A_m)  \\
   {\rm Im}(\pmb A_m) & {\rm Re}(\pmb A_m)  \\
\end{array}} \right]\in\mathbb{R}^{2N\times2N}
\end{equation}
and vector
\begin{equation}
\bar{\pmb x}=\left[ {\begin{array}{*{20}c}
   {\rm Re}(\pmb x)   \\
   {\rm Im}(\pmb x)  \\
\end{array}} \right]
\in\mathbb{R}^{2N}.
\end{equation}
Note that $\bar{\pmb A}_m$ is symmetric due to ${\rm Re}(\pmb A_m)={\rm Re}(\pmb A_m)^T$
and ${\rm Im}(\pmb A_m)=-{\rm Im}(\pmb A_m)^T$ because $\pmb A_m$ is Hermitian. It is also not
difficult to see $\pmb x^H\pmb A_m\pmb x =\bar{\pmb x}^T\bar{\pmb A}_m\bar{\pmb x}$.
Denoting the quadratic form as
\begin{equation}\label{qm:def}
    q_m(\bar{\pmb x})=\bar{\pmb x}^T\bar{\pmb A}_m\bar{\pmb x}
\end{equation}
we then rewrite \eqref{fm:def} as
\begin{equation}\label{fm:real}
    f_m(\bar{\pmb x}) = \left(q_m(\bar{\pmb x}) - b_m\right)^2
\end{equation}
and the original optimization problem of \eqref{Obj} becomes
\begin{equation}\label{Obj:real}
\min_{\bar{\pmb x}\in\mathbb{R}^{2N}} f(\bar{\pmb x})
:=\sum_{m=1}^{M}\left(q_m(\bar{\pmb x}) - b_m\right)^2.
\end{equation}
The objective function $f(\bar{\pmb x})$ is a multivariate
quartic polynomial of $\bar{\pmb x}=[\bar{x}_1,\cdots,\bar{x}_{2N}]^T$
since $q_m(\bar{\pmb x})$ is quadratic.
Minimizing multivariate fourth-order polynomial
is known to be NP-hard in general \cite{WF}.
In this work, we exploit the coordinate update strategy to minimize $f(\bar{\pmb x})$.
CD is an iterative procedure that successively minimizes the objective
function along coordinate directions. Denote the result of the $k$th
iteration as $\bar{\pmb x}^k=[\bar{x}_1^k,\cdots,\bar{x}_{2N}^k]^T$.
In the $k$th iteration, we minimize $f$ with respect to the
$i_k$th ($i_k\in\{1,\cdots,2N\}$) variable while keeping the remaining $2N-1$
variables $\{\bar{x}_i^k\}_{i\ne i_k}$ fixed. This is equivalent to
performing a one-dimensional search along the $i_k$th coordinate, which
can be expressed as
\begin{equation}\label{1D:min}
\alpha_k = \arg\min_{\alpha\in\mathbb{R}}
f\left(\bar{\pmb x}^k + \alpha\pmb e_{i_k}\right)
\end{equation}
where $\pmb e_{i_k}$ is the unit vector with the $i_k$th entry
being one and all other entries being zero. Then $\bar{\pmb x}$
is updated by
\begin{equation}\label{xk:update}
\bar{\pmb x}^{k+1} = \bar{\pmb x}^k + \alpha_k\pmb e_{i_k}
\end{equation}
which implies that only the $i_k$th component is updated:
\begin{equation}\label{xik:update}
\bar{x}_{i_k}^{k+1} \leftarrow \bar{x}_{i_k}^k + \alpha_k
\end{equation}
while other components remain unchanged. Since $\bar{\pmb x}^k$ is known,
$f\left(\bar{\pmb x}^k + \alpha\pmb e_{i_k}\right)$ is a univariate function
of $\alpha$. Thus, \eqref{1D:min} is a one-dimensional minimization problem. We will detail how to solve
it in the next subsection. Now one reason why we convert the complex-valued problem into real
is clear: this makes the scalar minimization problem of \eqref{1D:min} real-valued and easier to solve.
Otherwise, we still face a problem with a complex number, which in fact is a two-dimensional
optimization on the complex plane. The CD is outlined in Algorithm \ref{Algo:CD}.
\begin{algorithm}
    \caption{CD for Phase Retrieval}\label{Algo:CD}
    \algsetup{indent=2em}
    \begin{algorithmic}
    \vspace{1ex}
    \STATE \textbf{Initialization:} Choose $\bar{\pmb x}^0\in\mathbb{R}^{2N}$.
    \FOR{$k=0,1,\cdots,$}
    \STATE Choose index $i_k\in\{1,\cdots,2N\}$;
    \STATE $\alpha_k = \arg\min\limits_{\alpha\in\mathbb{R}}
    f\left(\bar{\pmb x}^k + \alpha\pmb e_{i_k}\right)$;
    \STATE $\bar{x}_{i_k}^{k+1} \leftarrow \bar{x}_{i_k}^k + \alpha_k$;
    \STATE \textbf{Stop} if termination condition is satisfied.
    \ENDFOR
    \end{algorithmic}
\end{algorithm}

There are several fashions to select the coordinate index $i_k$.
The following three selection rules are considered in this paper.
\begin{itemize}
  \item Cyclic rule: $i_k$ first takes 1, then 2 and so forth through $2N$.
  The process is then repeated starting with $i_k=1$ again. That is,
  $i_k$ takes value cyclically from $\{1,\cdots,2N\}$. Every $2N$ iterations
  are called one cycle or sweep. The cyclic rule is similar to the Gauss-Seidel iterative
  method for solving linear systems of equations \cite{Greenbaum}, where
  each coordinate is updated using a cyclic order.
  \item Random rule: $i_k$ is randomly selected from $\{1,\cdots,2N\}$ with equal probability.
  \item Greedy rule: $i_k$ is chosen as
\begin{equation}\label{GS:rule}
i_k= \arg\max_{i} |\nabla f_i(\bar{\pmb x}^k)|
\end{equation}
where
\begin{equation}\label{pder:def}
\nabla f_i(\bar{\pmb x}) = \frac{\partial f(\bar{\pmb x})}{\partial\bar{x}_i}
\end{equation}
is the partial derivative of $f(\bar{\pmb x})$ with respect to $\bar{x}_i$, i.e., the $i$th
component of the full gradient
\begin{equation}\label{grad:def}
\nabla f(\bar{\pmb x}) = \left[\frac{\partial f(\bar{\pmb x})}{\partial\bar{x}_1},
\cdots,\frac{\partial f(\bar{\pmb x})}{\partial\bar{x}_{2N}}\right]^T.
\end{equation}
The greedy rule is also called Gauss-Southwell rule \cite{YeBook}. Obviously, it chooses the coordinate with the largest (in absolute value)
partial derivative. Hence, computing the full gradient is required at each iteration while there is no need for the cyclic and
random rules. We refer the three CD methods with cyclic, random, and greedy rules to as CCD,
RCD, and GCD, respectively. It will be seen later that the GCD converges faster
than CCD and RCD at the expense of the extra full gradient calculation.
\end{itemize}

We call every $2N$ iterations of the CD as one cycle. Based on Wirtinger calculus \cite{WF},
the gradient of $f$ with respect to the complex vector $\pmb x$ is computed as
\begin{equation}\label{grad:complx}
\begin{aligned}
\nabla f(\pmb x) &= \frac{\partial f(\pmb x)}{\partial\pmb x^*}
=\frac{1}{2}\left(\frac{\partial f}{\partial\pmb x_R} + {\rm j}\frac{\partial f}{\partial\pmb x_I}\right)\\
&=2\sum_{m=1}^{M}\left(\left|\pmb a_m^H\pmb x\right|^2 - b_m\right) \pmb a_m\pmb a_m^H\pmb x
\end{aligned}
\end{equation}
with a complexity of $\mathcal{O}(MN)$. The gradient of $f$ with respect to
the real vector $\bar{\pmb x}$ is an expanded form of $\nabla f(\pmb x)$:
\begin{equation}\label{grad:real}
\nabla f(\bar{\pmb x})=\left[ {\begin{array}{*{20}c}
   {\rm Re}(\nabla f(\pmb x))   \\
   {\rm Im}(\nabla f(\pmb x))  \\
\end{array}} \right].
\end{equation}

\subsection{Closed-Form Solution of Coordinate Minimization}

The only remaining issue in the CD algorithm is on solving the scalar minimization
problem of \eqref{1D:min}. We now derive its closed-form solution as follows. For notational simplicity,
we omit the superscript and subscript $k$ in \eqref{1D:min}. That is, given the variable
$\bar{\pmb x}$ of the current iteration and the search direction $\pmb e_i$, we consider
minimizing the following univariate function
\begin{equation}\label{1D:min2}
\min_{\alpha\in\mathbb{R}} \varphi(\alpha):=f(\bar{\pmb x} + \alpha\pmb e_i).
\end{equation}
Employing \eqref{Obj:real}, $\varphi(\alpha)$ is expressed as
\begin{equation}\label{phi:alpha}
\varphi(\alpha)=\sum_{m=1}^{M}\left(q_m(\bar{\pmb x} + \alpha\pmb e_i) - b_m\right)^2
\end{equation}
where the $m$th term is
\begin{equation}\label{phi:m}
\varphi_m(\alpha)=\left(q_m(\bar{\pmb x} + \alpha\pmb e_i) - b_m\right)^2.
\end{equation}
We expand the quadratic function
\begin{equation}\label{qm:alpha}
\begin{aligned}
    q_m(\bar{\pmb x} + \alpha\pmb e_i)&= \alpha^2 \pmb e_i^T\bar{\pmb A}_m\pmb e_i
    + 2\alpha \pmb e_i^T\bar{\pmb A}_m\bar{\pmb x}
    + \bar{\pmb x}^T\bar{\pmb A}_m\bar{\pmb x}\\
    &{\buildrel\Delta \over = } ~c_{2,i}^m\alpha^2 + c_{1,i}^m\alpha + c_{0}^m
\end{aligned}
\end{equation}
where $c_{2,i}^m$, $c_{1,i}^m$, and $c_{0}^m$ are the coefficients of the
univariate quadratic polynomial. Note that the constant $c_{0}^m$ has no
relation to $i$. According to \eqref{Am:bar}, the coefficients of the quadratic term
can be simplified to
\begin{equation}\label{c2i}
\begin{aligned}
c_{2,i}^m &= \pmb e_i^T\bar{\pmb A}_m\pmb e_i
    = \left[\bar{\pmb A}_m\right]_{i,i}\\
    &=\left\{ \begin{array}{l}
 \left|[\pmb a_m]_i\right|^2,~i=1,\cdots,N \\
 \left|[\pmb a_m]_{i-N}\right|^2,~i=N+1,\cdots,2N. \\
 \end{array} \right.
\end{aligned}
\end{equation}
Using \eqref{Am:bar} and recalling $\pmb A_m=\pmb a_m\pmb a_m^H$, it is revealed that
\begin{equation}
\bar{\pmb A}_m\bar{\pmb x} =
\left[ {\begin{array}{*{20}c}
   {\rm Re}(\pmb A_m\pmb x)   \\
   {\rm Im}(\pmb A_m \pmb x)  \\
\end{array}} \right]
=
\left[ {\begin{array}{*{20}c}
   {\rm Re}\left(\left(\pmb a_m^H\pmb x\right) \pmb a_m\right)   \\
   {\rm Im}\left(\left(\pmb a_m^H \pmb x\right)\pmb a_m\right)  \\
\end{array}} \right].
\end{equation}
Hence, the coefficients of the linear term are computed as
\begin{equation}\label{c1i}
\begin{aligned}
c_{1,i}^m &=2\pmb e_i^T\bar{\pmb A}_m\bar{\pmb x}\\
& =\left\{ \begin{array}{l}
 {\rm Re}\left(\left(\pmb a_m^H\pmb x\right) [\pmb a_m]_i\right),~i=1,\cdots,N \\
 {\rm Im}\left(\left(\pmb a_m^H \pmb x\right)[\pmb a_m]_{i-N}\right),~i=N+1,\cdots,2N. \\
 \end{array} \right.
\end{aligned}
\end{equation}
The constant term is
\begin{equation}\label{c0i}
    c_{0}^m = \bar{\pmb x}^T\bar{\pmb A}_m\bar{\pmb x}
    = \pmb x^H\pmb A_m\pmb x =\left|\pmb a_m^H\pmb x\right|^2.
\end{equation}
Since $q_m(\bar{\pmb x} + \alpha\pmb e_i)$ is quadratic,
$\varphi_m(\alpha)$ of \eqref{phi:m} is a univariate quartic polynomial of $\alpha$,
which is expressed as
\begin{equation}\label{phim:alpha}
    \varphi_m(\alpha)
    =d_{4,i}^m\alpha^4 + d_{3,i}^m\alpha^3 + d_{2,i}^m\alpha^2 + d_{1,i}^m\alpha + d_{0}^m
\end{equation}
where $\{d_{j,i}^m\}_{j=1}^4$ and $d_{0}^m$ are the coefficients of the polynomial.
Note that $d_{0}^m$ is not related to $i$. Plugging \eqref{qm:alpha} into \eqref{phi:m}, we obtain
\begin{equation}\label{dm}
\begin{aligned}
    &d_{4,i}^m = \left(c_{2,i}^m\right)^2\\
    &d_{3,i}^m = 2c_{2,i}^m c_{1,i}^m\\
    &d_{2,i}^m = \left(c_{1,i}^m\right)^2 + 2c_{2,i}^m\left(c_{0}^m - b_m\right)\\
    &d_{1,i}^m = 2c_{1,i}^m\left(c_{0}^m - b_m\right)\\
    &d_{0}^m = \left(c_{0}^m - b_m\right)^2.
\end{aligned}
\end{equation}
Since $\varphi(\alpha)= \sum_{m=1}^M \varphi_m(\alpha)$, it is clear that
the coefficients of the quartic polynomial
\begin{equation}\label{Obj:alpha}
    \varphi(\alpha) = d_{4,i} \alpha^4 + d_{3,i} \alpha^3
    + d_{2,i}\alpha^2 + d_{1,i}\alpha + d_0
\end{equation}
correspond to the sums of those of $\{\varphi_m(\alpha)\}_{m=1}^M$, i.e.,
\begin{equation}\label{coef:d}
    d_0 = \sum_{m=1}^M d_{0}^m, \quad d_{j,i} = \sum_{m=1}^M d_{j,i}^m, \quad j = 1,\cdots,4.
\end{equation}
The minimum point of $\varphi(\alpha)$ must be one of stationary points, i.e., the roots of the derivative
\begin{equation}\label{dphi:alpha}
   \varphi'(\alpha)= 4d_{4,i} \alpha^3 + 3d_{3,i} \alpha^2
    + 2d_{2,i}\alpha + d_{1,i} = 0.
\end{equation}
Equation \eqref{dphi:alpha} refers to finding the roots of a univariate cubic polynomial,
which is easy and fast because there is a closed-form
solution \cite{Cubic:Root2}. Since the coefficients of the cubic equation
are real-valued, there are only two possible cases on the roots.
The first case is that \eqref{dphi:alpha} has a real root and a pair of complex conjugate roots.
In this case, the minimizer is the unique real root because the optimal solution of a real-valued
problem must be real-valued. The second case is that \eqref{dphi:alpha} has three real roots. Then the optimal
$\alpha$ is the real root associated with the minimum objective. Once the coefficients of \eqref{dphi:alpha}
are obtained, the complexity of calculating the roots of a cubic polynomial is merely $\mathcal{O}(1)$.
Herein, the second reason why we recast the complex-valued problem into real is clear: by this fashion,
it results in root finding of a cubic equation with real coefficients, which has a closed-form
solution and is much simpler than the case with complex coefficients.

\emph{Computational Complexity:} The leading computational cost at each
iteration of the CD is calculating the coefficients
$\{d_{j,i}\}_{j=1}^4$, or equivalently, computing $c_{2,i}^m$, $c_{1,i}^m$,
and $c_{0}^m$ with $m=1,\cdots,M$.\footnote{This is because $\{d_{j,i}\}_{j=1}^4$ can be easily calculated from $c_{2,i}^m$, $c_{1,i}^m$,
and $c_{0}^m$ according to \eqref{dm} and \eqref{coef:d}.} From \eqref{c2i}, $c_{2,i}^m$ is
just the squared modulus of $[\pmb a_m]_i$ and can be pre-computed in advance
before iteration, which requires $\mathcal{O}(M)$ multiplications for
determining all $M$ coefficients $\{c_{2,i}^m\}_{m=1}^M$. According to
\eqref{c1i} and \eqref{c0i}, we need to compute $\{\pmb a_m^H\pmb x\}_{m=1}^M$
in order to obtain $\{c_{1,i}^m\}_{m=1}^M$ and $\{c_{0}^m\}_{m=1}^M$. This
involves a matrix-vector multiplication $\pmb A\pmb x$, where the $m$th row of
the matrix $\pmb A$ is $\pmb a_m^H$, i.e.,
\begin{equation}
\pmb A=\left[ {\begin{array}{*{20}c}
   \pmb a_1^H   \\
   \vdots\\
   \pmb a_M^H  \\
\end{array}} \right]\in \mathbb{C}^{M\times N}.
\end{equation}
At first glance,
the matrix-vector multiplication requires a complexity of $\mathcal{O}(MN)$.
However, this complexity can be reduced to $\mathcal{O}(M)$ per iteration for the CD.
By observing \eqref{1D:min}, we know that only one single element changes in two
consecutive iterations. Specifically, we have
\begin{equation}
   \pmb x^{k+1} - \pmb x^k = (x_j^{k+1} - x_j^k)\pmb e_j,
   \quad j = \left\{ \begin{array}{l}
 i_k, ~{\rm if}~ j\le N \\
 i_k - N, ~{\rm otherwise}\\
 \end{array} \right.
\end{equation}
which yields
\begin{equation}
   \pmb A\pmb x^{k+1} = \pmb A\pmb x^k + (x_j^{k+1} - x_j^k)\pmb A_{:,j}
\end{equation}
where $\pmb A_{:,j}$ represents the $j$th column of $\pmb A$. It is only required to
compute $\pmb A\pmb x^0$ before iteration. After that, this matrix-vector
product can be efficiently updated from that of the previous
iteration by a cheap computation of a scalar-vector multiplication $(x_j^{k+1} - x_j^k)\pmb A_{:,j}$, which merely costs
$\mathcal{O}(M)$ operations. In summary, the complexity of CCD and RCD is
$\mathcal{O}(M)$ per iteration. Therefore, the complexity of $2N$ iterations,
i.e., a cycle for CCD, is the same as that of the WF method using full gradient descent.
While for GCD, an extra cost for computing the full gradient is needed, which results in
a complexity of $\mathcal{O}(MN)$.

\emph{Initialization and Termination:} The spectral method in \cite{WF} provides
a good initial value for phase retrieval. For Gaussian measurement model
and in the absence of noise, we have \cite{WF}
\begin{equation}
   \mathbb{E}\left[\frac{1}{M}\sum_{m=1}^M b_m \pmb a_m\pmb a_m^H\right]
    = \pmb I + 2\pmb x\pmb x^H.
\end{equation}
Since $\pmb x$ is the principal eigenvector of $\pmb I + 2\pmb x\pmb x^H$
associated with the largest eigenvalue, the principal eigenvector\footnote{The squared norm of the eigenvector
is set to $(N\|\pmb b\|_1)/(\sum_m\|\pmb a_m\|^2)$.} of the matrix
$\frac{1}{M}\sum_{m=1}^M b_m \pmb a_m\pmb a_m^H$, which is an estimate of $\pmb I + 2\pmb x\pmb x^H$,
is taken as the initial value $\pmb x^0$. More details of the spectral method for initialization
can be found in \cite{WF}. There are several measures for terminating the CD algorithm.
For example, the reduction of the objective function can be used to check for convergence.
Specifically, the iteration is terminated when
\begin{equation}\label{Stop:rule}
f(\bar{\pmb x}^k)-f(\bar{\pmb x}^{k+1})<\texttt{TOL}
\end{equation}
holds, where $\texttt{TOL}>0$ is a small tolerance parameter. Note that the CD
monotonically decreases the objective function, implying
$f(\bar{\pmb x}^k)-f(\bar{\pmb x}^{k+1})>0$.

\section{Convergence Analysis}\label{Sec:Converge}

Most existing convergence analyses for CD assume that the objective function is convex and
the gradient is Lipschitz continuous \cite{Wright,Nesterov,Beck}. However, the objective function for phase retrieval is
quartic and hence nonconvex. As shown in \eqref{grad:complx} and \eqref{grad:real}, the gradient is not
Lipschitz continuous. Therefore, the available convergence analyses are not applicable to
the CD for phase retrieval. In this section, We first prove that the three CD algorithms globally converge to a stationary point from any initial value. Then, it is proved that the sequence of the iterates generated by the RCD locally converges to the global minimum point in expectation at a geometric rate under a mild assumption. This implies that in the absence of noise, the RCD achieves exact phase retrieval under a moderate condition.

\subsection{Global Convergence to Stationary Point}

We first present two lemmas used in the proof.

\emph{Lemma 1:} Given any finite initial value $\bar{\pmb x}^0\in \mathbb{R}^{2N}$ and $f(\bar{\pmb x}^0)=f_0$, the
sublevel set of $f(\bar{\pmb x})$
\begin{equation}\label{Level:set}
\mathcal{S}_{f_0} = \{\bar{\pmb x} | f(\bar{\pmb x})\le f_0\}
\end{equation}
is compact, viz. bounded and closed. The iterates of the three CD algorithms, i.e.,
$\bar{\pmb x}^k$, $k=0,1,\cdots$, are in the compact set $\mathcal{S}_{f_0}$.

\emph{Proof:} If $\|\bar{\pmb x}\|\rightarrow \infty$, then $f(\bar{\pmb x})\rightarrow \infty$
since $f(\bar{\pmb x})$ is quartic. The converse-negative proposition implies that
$f(\bar{\pmb x})\le f_0<\infty$ guaranteeing $\|\bar{\pmb x}\|<\infty$ for all
$\bar{\pmb x} \in \mathcal{S}_{f_0}$. Hence, $\mathcal{S}_{f_0}$ is bounded.
Now it is clear that all the points in the sublevel set satisfy
$f(\bar{\pmb x})\in[0, f_0]$ as we also have $f(\bar{\pmb x})\ge 0$. Since the mapping $f(\bar{\pmb x})$ is continuous
and the image $[0,f_0]$ is a closed set, the inverse image
$\{\bar{\pmb x} | 0\le f(\bar{\pmb x})\le f_0\}$ is also closed.
This completes the proof that $\mathcal{S}_{f_0}$ is compact. The CD
monotonically decreases $f(\bar{\pmb x})$, meaning that $f(\bar{\pmb x}^k)
\le f(\bar{\pmb x}^{k-1})\le \cdots \le f(\bar{\pmb x}^0)= f_0$.
Therefore, all the iterates $\{\bar{\pmb x}^k\}$ must be in the compact
set $\mathcal{S}_{f_0}$. \hfill $\Box$

Lemma 1 guarantees that we can limit the analysis in the compact
set $\mathcal{S}_{f_0}$ rather than the whole domain of $f(\bar{\pmb x})$. In the following, Lemma 2 states that the partial derivatives of $f(\bar{\pmb x})$ are \emph{locally} Lipschitz
continuous on $\mathcal{S}_{f_0}$, although they are not \emph{globally}
Lipschitz continuous over the whole domain $\mathbb{R}^{2N}$.

\emph{Lemma 2:} On the compact set $\mathcal{S}_{f_0}$, the gradient
$\nabla f(\bar{\pmb x})$ is \emph{component-wise} Lipschitz continuous.
That is, for each $i=1,\cdots,2N$, we have
\begin{equation}\label{Comp:Lipsch}
|\nabla_i f(\bar{\pmb x} + t\pmb e_i) -\nabla_i f(\bar{\pmb x})|\le L_i|t|,~t\in \mathbb{R}
\end{equation}
for all $\bar{\pmb x}, \bar{\pmb x} + t\pmb e_i \in \mathcal{S}_{f_0}$, where $L_i>0$
is referred to as the component-wise Lipschitz constant on $\mathcal{S}_{f_0}$.
Further, it follows
\begin{equation}\label{CLipsch2}
f(\bar{\pmb x} + t\pmb e_i) \le f(\bar{\pmb x}) + t\nabla_i f(\bar{\pmb x})
+ \frac{L_i}{2}t^2.
\end{equation}

\emph{Proof:} For $t=0$, both sides of \eqref{Comp:Lipsch} are equal to 0
and \eqref{Comp:Lipsch} holds. For $t\ne0$, it means that
$\|\bar{\pmb x} + t\pmb e_i -\bar{\pmb x}\| =t \ne0$. Since the function
\begin{equation}
\frac{|\nabla_i f(\bar{\pmb x} + t\pmb e_i) -\nabla_i f(\bar{\pmb x})|}
{\|\bar{\pmb x} + t\pmb e_i -\bar{\pmb x}\|}
\end{equation}
is continuous on the compact set $\mathcal{S}_{f_0}$, its minimum over
$\mathcal{S}_{f_0}$, namely, $L_i$, is attained by
Weierstrass' theorem \cite{Royden}. Then
\begin{equation}
\max \frac{|\nabla_i f(\bar{\pmb x} + t\pmb e_i) -\nabla_i f(\bar{\pmb x})|}
{\|\bar{\pmb x} + t\pmb e_i -\bar{\pmb x}\|} = L_i
\end{equation}
immediately elicits \eqref{Comp:Lipsch}. The following second-order partial derivative
\begin{equation}\label{SO-Derivative}
\nabla_{i,i}^2 f(\bar{\pmb x}) = \frac{\partial^2 f(\bar{\pmb x})}{\partial \bar{x}_i^2}.
\end{equation}
is well defined since $f(\bar{\pmb x})$ is twice continuously differentiable, which represents the $(i,i)$ entry of the Hessian matrix:
\begin{equation}\label{Hessian:def}
\nabla^2 f(\bar{\pmb x}) = \frac{\partial^2 f(\bar{\pmb x})}
{\partial \bar{\pmb x}\partial \bar{\pmb x}^T} \in \mathbb{R}^{2N\times2N}.
\end{equation}
Noting that $\nabla_{i,i}^2 f(\bar{\pmb x})$ is the partial derivative of $\nabla_i f(\bar{\pmb x})$
with respect to $\bar{x}_i$ and by \eqref{Comp:Lipsch}, we obtain
\begin{equation}\label{Hii:Li}
\nabla_{i,i}^2 f(\bar{\pmb x}) = \lim_{t\rightarrow 0}
\frac{\nabla_i f(\bar{\pmb x} + t\pmb e_i) -\nabla_i f(\bar{\pmb x})}{t}\le L_i
\end{equation}
which holds for all $\bar{\pmb x}\in\mathcal{S}_{f_0}$.
Applying Taylor's theorem and \eqref{Hii:Li}, there exists a $\gamma\in[0,1]$ with
$\bar{\pmb x}+\gamma t\pmb e_i\in\mathcal{S}_{f_0}$ such that
\begin{equation}\label{Taylor}
\begin{aligned}
f(\bar{\pmb x} + t\pmb e_i) &= f(\bar{\pmb x}) + t\nabla f(\bar{\pmb x})^T\pmb e_i
+ \frac{t^2}{2}\pmb e_i^T \nabla^2 f(\bar{\pmb x}+\gamma t\pmb e_i )\pmb e_i \\
&=f(\bar{\pmb x}) + t\nabla f_i(\bar{\pmb x})
+ \frac{t^2}{2}\nabla_{i,i}^2 f(\bar{\pmb x}+\gamma t\pmb e_i )\\
&\le f(\bar{\pmb x}) + t\nabla f_i(\bar{\pmb x})
+ \frac{L_i}{2}t^2.
\end{aligned}
\end{equation}
\hfill $\Box$

The component-wise Lipschitz constant $L_i$ is not easy to compute or estimate
because the partial derivatives are complicated multivariate polynomials.
However, our CD algorithms do not require $L_i$. This quantity is just used
for theoretical convergence analysis. The minimum and maximum of all the component-wise
Lipschitz constants, respectively, are:
\begin{equation}
L_{\min} = \min_{1\le i\le 2N} L_i, \quad L_{\max} = \max_{1\le i\le 2N} L_i.
\end{equation}
Employing similar steps of Lemma 2, we can prove that the full gradient
$\nabla f(\bar{\pmb x})$ is Lipschitz continuous:
\begin{equation}\label{full:Lipsch}
\|\nabla f(\bar{\pmb x}) -\nabla f(\bar{\pmb z})\|
\le L\|\bar{\pmb x} - \bar{\pmb z}\|
\end{equation}
with $L$ being the ``full'' Lipschitz constant. It is not difficult to show $L\le \sum_i L_i$
and thus we further have $L\le 2N L_{\max}$.

\emph{Theorem 1:} The CCD, RCD, and GCD globally converge to a stationary point of the
multivariate quartic polynomial from an arbitrary initialization.

\emph{Proof:} Based on the component-wise Lipschitz continuous property of \eqref{CLipsch2},
it is derived that:
\begin{equation}
\begin{aligned}
f(\bar{\pmb x}^{k+1}) &= \min_{\alpha}f(\bar{\pmb x}^k + \alpha\pmb e_{i_k})\\
&\le f(\bar{\pmb x}^k + \alpha\pmb e_{i_k})|_{\alpha = -\nabla_{i_k} f(\bar{\pmb x}^k)/L_{i_k}}\\
&= f\left(\bar{\pmb x}^k -\frac{\nabla_{i_k} f(\bar{\pmb x}^k)}{L_{i_k}} \pmb e_{i_k}\right)\\
&\le f(\bar{\pmb x}^k) - \frac{(\nabla f_{i_k}(\bar{\pmb x}^k))^2}{L_{i_k}}
+ \frac{L_{i_k}}{2}\frac{(\nabla f_{i_k}(\bar{\pmb x}^k))^2}{L_{i_k}^2}\\
&= f(\bar{\pmb x}^k) - \frac{1}{2L_{i_k}}(\nabla f_{i_k}(\bar{\pmb x}^k))^2\\
&\le f(\bar{\pmb x}^k) - \frac{1}{2L_{\max}}(\nabla f_{i_k}(\bar{\pmb x}^k))^2
\end{aligned}
\end{equation}
from which we obtain a lower bound on the progress made by each CD iteration
\begin{equation}\label{Basic:LB}
f(\bar{\pmb x}^k) - f(\bar{\pmb x}^{k+1}) \ge \frac{1}{2L_{\max}}(\nabla f_{i_k}(\bar{\pmb x}^k))^2.
\end{equation}
For different rules of index selection, the right-hand side of \eqref{Basic:LB} will differ.
We discuss the GCD, RCD, and CCD, one by one as follows. For GCD, it chooses the index with the
largest partial derivative in magnitude. With the use of \eqref{GS:rule}, we then have:
\begin{equation} \label{so1}
(\nabla f_{i_k}(\bar{\pmb x}^k))^2 = \|\nabla f(\bar{\pmb x}^k)\|_{\infty}^2
\ge \frac{1}{2N}\|\nabla f(\bar{\pmb x}^k)\|^2.
\end{equation}
Substituting \eqref{so1} into \eqref{Basic:LB} leads to the following lower
bound of the progress of one GCD iteration
\begin{equation}\label{GCD:LB}
f(\bar{\pmb x}^k) - f(\bar{\pmb x}^{k+1})
\ge \frac{1}{4NL_{\max}}\|\nabla f(\bar{\pmb x}^k)\|^2.
\end{equation}
This means that one GCD iteration decreases the objective function
with an amount of at least $\frac{\|\nabla f(\bar{\pmb x}^k)\|^2}{4NL_{\max}}$.
Setting $k=0, \cdots, j$, in \eqref{GCD:LB} and summing over all inequalities yields
\begin{equation}\label{GCD:LB2}
\sum_{k=0}^j\|\nabla f(\bar{\pmb x}^k)\|^2
\le 4N L_{\max}\left(f(\bar{\pmb x}^0) - f(\bar{\pmb x}^{j+1})\right)\le 4N L_{\max} f_0
\end{equation}
where we use $f(\bar{\pmb x}^{j+1})\ge 0$. Taking the limit as $j\rightarrow \infty$
on \eqref{GCD:LB2}, we get a convergent series
\begin{equation}\label{GCD:LB3}
\sum_{k=0}^{\infty}\|\nabla f(\bar{\pmb x}^k)\|^2\le 4N L_{\max} f_0.
\end{equation}
If a series converges, then its terms approach to zero, which indicates
\begin{equation}\label{}
\lim_{k\rightarrow \infty}\nabla f(\bar{\pmb x}^k)=\pmb 0
\end{equation}
i.e., the GCD converges to a stationary point.

For RCD, since $i_k$ is a random variable, $f(\bar{\pmb x}^{k+1})$ is also random and we consider its expected value:
\begin{equation}\label{RCD:Expect}
\begin{aligned}
\mathbb{E}\left[f(\bar{\pmb x}^{k+1})\right] &\le \mathbb{E}\left[f(\bar{\pmb x}^k) -
\frac{1}{2L_{\max}}(\nabla f_{i_k}(\bar{\pmb x}^k))^2 \right]\\
&= f(\bar{\pmb x}^k)-\frac{1}{2L_{\max}}\sum_{i=1}^{2N}\frac{1}{2N}(\nabla f_{i}(\bar{\pmb x}^k))^2\\
&= f(\bar{\pmb x}^k) -\frac{1}{4N L_{\max}}\|\nabla f(\bar{\pmb x}^k)\|^2
\end{aligned}
\end{equation}
where the fact that $i_k$ is uniformly sampled from $\{1,\cdots,2N\}$
with equal probability of $1/(2N)$ is employed. Then the RCD at least obtains a reduction
on the objective function in expectation
\begin{equation}\label{RCD:LB}
f(\bar{\pmb x}^k)- \mathbb{E}\left[f(\bar{\pmb x}^{k+1})\right]
\ge \frac{1}{4N L_{\max}}\|\nabla f(\bar{\pmb x}^k)\|^2.
\end{equation}
Following similar steps in the GCD, it is easy to prove that the expected gradient
of the RCD approaches to the zero vector and thus it converges to a stationary point in expectation.
For CCD, $i_k$ takes value cyclically from $\{1,\cdots,2N\}$. Applying Lemma 3.3
in \cite{Beck} for cyclic block CD, we can derive a lower bound of the decrease of the objective function
after $2N$ CCD iterations:
\begin{equation}\label{CCD:LB}
f(\bar{\pmb x}^k) - f(\bar{\pmb x}^{k+2N})
\ge \frac{\|\nabla f(\bar{\pmb x}^k)\|^2}{4L_{\max}(1+2N L^2/L_{\min}^2)}.
\end{equation}
Setting $k=0,2N,\cdots,2jN$, in \eqref{CCD:LB} and summing over all the inequalities yields
\begin{equation}\label{CCD:LB2}
\begin{aligned}
\sum_{k=0}^j\|\nabla f(\bar{\pmb x}^{2kN})\|^2
\le \frac{\left(f(\bar{\pmb x}^0) - f\left(\bar{\pmb x}^{2(j+1)N}\right)\right)}
{4L_{\max}(1+2N L^2/L_{\min}^2)}\\
\le \frac{f(\bar{\pmb x}^0)}
{4L_{\max}(1+2N L^2/L_{\min}^2)}.
\end{aligned}
\end{equation}
Taking the limit as $j\rightarrow \infty$ of \eqref{CCD:LB2} yields a convergent series.
Thus, the gradient approaches to zero, indicating that the CCD converges to
a stationary point. \hfill $\Box$

We emphasize that the ``global'' convergence to a stationary point means that CD converges
from an arbitrary initial value. Unlike local convergence, it does not require the initial
value to be close enough to the stationary point.

\emph{Remark 1:} Several existing convergence analyses of (block) CD, e.g., Proposition 2.7.1
of Bertsekas' book \cite{Bertsekas} and page 153 of \cite{YeBook},
assume that the minimum of each block/coordinate is uniquely attained. However, our
analysis in Theorem 1 does not require this assumption.

\emph{Remark 2:} Theorem 6.1 of \cite{Luke} provides a convergence result
for a descent method using update formula $\bar{\pmb x}^{k+1} = \bar{\pmb x}^k + \delta_k\pmb t^k$,
where $\pmb t^k$ is a descent direction and $\delta_k>0$ is the stepsize.
Theorem 6.1 of \cite{Luke} has proved
\begin{equation}\label{Luke:Result}
    \left\langle \nabla f(\bar{\pmb x}^k),\pmb t^k \right\rangle \rightarrow 0
\end{equation}
if $\delta_k>0$ is determined by an inexact line search procedure to ensure
sufficient decrease at each iteration. Since the full gradient descent
method adopts $\pmb t^k = -\nabla f(\bar{\pmb x}^k)$, \eqref{Luke:Result}
becomes $\|f(\bar{\pmb x}^k)\| \rightarrow 0$ and hence
$f(\bar{\pmb x}^k) \rightarrow \pmb 0$. Then Theorem 6.1 of \cite{Luke} proves that the full gradient descent
converges to a stationary point. For CDs, it has $\pmb t^k = \pmb e_{i_k}$ and
\eqref{Luke:Result} becomes $\nabla_{i_k} f(\bar{\pmb x}^k) \rightarrow 0$.
Clearly, we can only conclude a single partial derivative approaches zero
and cannot conclude other partial derivatives approach zero. Therefore, Theorem 6.1
of \cite{Luke} cannot be used to prove the convergence to a stationary point for CDs.
Moreover, the proof of Theorem 6.1 of \cite{Luke} requires the gradient
is globally Lipschitz continuous, which results in that it is not applicable to our problem.
In addition, \cite{Luke} uses an inexact line search for stepsize while the CDs
adopt exact coordinate minimization. The self-contained convergence analysis of CDs
is totally different from \cite{Luke}.

\emph{Remark 3:} Even when there are enough samples, the Hessian matrix $\nabla^2 f(\bar{\pmb x})$
close to the minimizer $\bar{\pmb x}^{\star}$ has $2N-1$ positive eigenvalues, and
the remaining eigenvalue can be zero, positive, or negative. This implies that $f(\bar{\pmb x})$
can never be locally convex no matter how small the local region around $\bar{\pmb x}^{\star}$ is.
Therefore, the established results \cite{Wright,Nesterov,Beck} for convergence rate
using convexity are not applicable for our nonconvex problem.

\subsection{Local Convergence to Global Minimum}

Theorem 1 just shows that the CD algorithm converges to a stationary point.
A further question is: can the CD converge to the global minimizer and hence exactly
recovers the original signal? At first glance, it seems impossible because even finding
a local minimum of a fourth-order polynomial is known to be NP-hard in general \cite{WF,NP:hard}.
However, the answer is yes under the condition that the sample size is large enough.
The backbone of the proof is based on a statistical analysis of the gradient
of the nonconvex objective function established by Cand\`{e}s \emph{et al.} \cite{WF}.
It is worth mentioning that the convergence analysis of WF \cite{WF} is for the complex-valued full gradient method
and cannot be directly applied to our real-valued problem using coordinate minimization.

Recall that if $\pmb x^{\star}$ is an optimal solution of \eqref{Obj}, then all the
elements of the following set
\begin{equation}\label{Pc:set}
\mathcal{P}_c:=\left\{{\rm e}^{{\rm j}\phi}\pmb x^{\star}
,~\phi\in[0,2\pi)\right\}
\end{equation}
are also optimal solutions of \eqref{Obj}. The distance of a vector $\pmb z\in \mathbb{C}^N$
to $\mathcal{P}_c$ is defined as:
\begin{equation}\label{dist:cplx}
{\rm dist}(\pmb z,\mathcal{P}_c)
=\min_{\phi}\|\pmb z - {\rm e}^{{\rm j}\phi}\pmb x^{\star} \|
\end{equation}
and the minimum of \eqref{dist:cplx} attains at $\phi=\phi(\pmb z)$.
Similarly, the set of all optimal solutions of the real-valued problem \eqref{Obj:real}
is defined as
\begin{equation}\label{P:set}
\mathcal{P}:=\left\{\left[ {\begin{array}{*{20}{c}}
   {\rm Re}({\rm e}^{{\rm j}\phi}\pmb x^{\star})  \\
   {\rm Im}({\rm e}^{{\rm j}\phi}\pmb x^{\star}) \\
\end{array}} \right]{\buildrel \Delta \over = }~T_{\phi}(\bar{\pmb x}^{\star})
,~\phi\in[0,2\pi)\right\}
\end{equation}
where $\bar{\pmb x}^{\star} =[{\rm Re}(\pmb x^{\star})^T, {\rm Im}(\pmb x^{\star})^T]^T$ is a global minimizer
of \eqref{Obj:real}. That is, $T_{\phi}(\bar{\pmb x}^{\star})$ denotes the effect of
a phase rotation to $\bar{\pmb x}^{\star}$. The projection of $\bar{\pmb x}^k$
onto $\mathcal{P}$ is the point in $\mathcal{P}$ closest to $\bar{\pmb x}^k$, which
is denoted as $T_{\phi_k}(\bar{\pmb x}^{\star})$ where
\begin{equation}\label{Proj2P}
\phi_k = \arg\min_{\phi} \|\bar{\pmb x}^k - T_{\phi}(\bar{\pmb x}^{\star})\|.
\end{equation}
Then the distance of $\bar{\pmb x}^k$ to $\mathcal{P}$ is
\begin{equation}\label{dist}
\begin{aligned}
{\rm dist}(\bar{\pmb x}^k,\mathcal{P})
&=\min_{\phi}\|\bar{\pmb x}^k - T_{\phi}(\bar{\pmb x}^{\star}) \|\\
&=\|\bar{\pmb x}^k - T_{\phi_k}(\bar{\pmb x}^{\star}) \|.
\end{aligned}
\end{equation}
Our goal is to prove ${\rm dist}(\bar{\pmb x}^k,\mathcal{P})\rightarrow 0$.
The following lemma of \cite{WF}, which essentially states that the gradient of
the objective function is well behaved, is crucial to our proof.

\emph{Lemma 3:} For any $\pmb z\in \mathbb{C}^N$ with
${\rm dist}(\pmb z,\mathcal{P}_c)\le \epsilon$, the \emph{regularity condition}
\begin{equation}\label{Reg:Cond}
{\rm Re}\left(\left\langle \nabla f(\pmb z),
\pmb z-{\rm e}^{{\rm j}\phi(\pmb z)}\pmb x^{\star}\right\rangle\right)
\ge \rho~{\rm dist}(\pmb z,\mathcal{P}_c) + \eta \|\nabla f(\pmb z)\|^2
\end{equation}
where $\rho>0$ and $\eta>0$, holds with high probability if the
number of measurements satisfies $M\ge C_0 N\log N$ with $C_0>0$ being a
sufficiently large constant.

The detailed proof of Lemma 3 can be found in Condition 7.9, Theorem 3.3, and Sections 7.5--7.7 of \cite{WF}.

Although the regularity condition of Lemma 3 corresponds to the complex-valued case, we at once obtain
the real-valued version according to \eqref{grad:real}, \eqref{P:set}, and \eqref{Proj2P}.
For $\bar{\pmb x}^k$ satisfying ${\rm dist}(\bar{\pmb x}^k,\mathcal{P})\le \epsilon$, we have
\begin{equation}\label{RegCond:Real}
\left\langle \nabla f(\bar{\pmb x}^k),
\bar{\pmb x}^k-T_{\phi_k}(\bar{\pmb x}^{\star})\right\rangle
\ge \rho~{\rm dist}(\bar{\pmb x}^k,\mathcal{P}) + \eta \|\nabla f(\bar{\pmb x}^k)\|^2.
\end{equation}

\emph{Theorem 2:} Assume that the sample size satisfies $M\ge C_0 N\log N$
with a sufficiently large $C_0$ and ${\rm dist}(\bar{\pmb x}^0,\mathcal{P})\le\epsilon$.
The iterates of the RCD with a slight modification,
in which the one-dimensional search is limited to a line segment, i.e.,
\begin{equation}\label{alphak:min}
\alpha_k = \arg\min_{\alpha}
f\left(\bar{\pmb x}^k + \alpha\pmb e_{i_k}\right),
~{\rm s.t.}~|\alpha| \le 2\eta |\nabla f_{i_k}(\bar{\pmb x}^k)|
\end{equation}
satisfy ${\rm dist}(\bar{\pmb x}^k,\mathcal{P})\le \epsilon$ for all $k$ and converge
to $\mathcal{P}$ in expectation with high probability at a geometric rate\footnote{The
geometric convergence rate is also called linear convergence rate in the optimization
literature. It indicates that the logarithm of the error decreases linearly.}
\begin{equation}\label{Geometric:Conv}
\mathbb{E}\left[{\rm dist}^2(\bar{\pmb x}^{k+1},\mathcal{P})\right] \le
\left(1-\frac{\rho\gamma_{\min}}{N}\right)^k{\rm dist}^2(\bar{\pmb x}^0,\mathcal{P})
\end{equation}
where $\gamma_{\min}>0$.

\emph{Proof:} The updating equation of the CD, i.e., $\bar{\pmb x}^{k+1} = \bar{\pmb x}^k + \alpha_k\pmb e_{i_k}$,
is equivalently expressed as
\begin{equation}\label{xk:gamma}
\bar{\pmb x}^{k+1} = \bar{\pmb x}^k - \gamma_k\nabla f_{i_k}(\bar{\pmb x}^k)\pmb e_{i_k}
\end{equation}
where $\gamma_k= -\alpha_k/\nabla f_{i_k}(\bar{\pmb x}^k)$. It requires $\gamma_k>0$ to
ensure $f(\bar{\pmb x}^{k+1}) < f(\bar{\pmb x}^k)$. Hence, $|\alpha_k|\le 2\eta |\nabla f_i(\bar{\pmb x}^k)|$
means $0<\gamma_k\le2\eta$. Employing the development starting from \eqref{dist}, it follows
\begin{equation}\label{dist:progress}
\begin{aligned}
{\rm dist}^2(\bar{\pmb x}^{k+1},\mathcal{P})
&=\|\bar{\pmb x}^{k+1} - T_{\phi_{k+1}}(\bar{\pmb x}^{\star})\|^2\\
&\le \|\bar{\pmb x}^{k+1} - T_{\phi_k}(\bar{\pmb x}^{\star}) \|^2\\
&=\|\bar{\pmb x}^k - \gamma_k\nabla f_i(\bar{\pmb x}^k)\pmb e_{i_k} - T_{\phi_k}(\bar{\pmb x}^{\star}) \|^2\\
&=\|\bar{\pmb x}^k - T_{\phi_k}(\bar{\pmb x}^{\star}) \|^2+ \gamma_k^2(\nabla f_{i_k}(\bar{\pmb x}^k))^2\\
&\quad~ - 2\gamma_k\nabla f_{i_k}(\bar{\pmb x}^k)\pmb e_{i_k}^T \left(\bar{\pmb x}^k - T_{\phi_k}(\bar{\pmb x}^{\star})\right)\\
&={\rm dist}^2(\bar{\pmb x}^k,\mathcal{P}) + \gamma_k^2(\nabla f_{i_k}(\bar{\pmb x}^k))^2\\
&\quad~- 2\gamma_k\nabla f_{i_k}(\bar{\pmb x}^k)\left[\bar{\pmb x}^k - T_{\phi_k}(\bar{\pmb x}^{\star})\right]_{i_k}.
\end{aligned}
\end{equation}
It is already known that $\mathbb{E}\left[(\nabla f_{i_k}(\bar{\pmb x}^k))^2\right]
=\|\nabla f(\bar{\pmb x}^k)\|^2/(2N)$ by \eqref{RCD:Expect}. We also have
\begin{equation}\label{InnerProd:Expe}
\begin{aligned}
&\mathbb{E}\left[ \nabla f_{i_k}(\bar{\pmb x}^k)
\left[\bar{\pmb x}^k- T_{\phi_k}(\bar{\pmb x}^{\star})\right]_{i_k} \right] \\
&= \frac{1}{2N}\sum_{i=1}^{2N} \nabla f_{i}(\bar{\pmb x}^k)
\left[\bar{\pmb x}^k- T_{\phi_k}(\bar{\pmb x}^{\star})\right]_{i}\\
&= \frac{1}{2N} \left\langle \nabla f(\bar{\pmb x}^k),
\bar{\pmb x}^k-T_{\phi_k}(\bar{\pmb x}^{\star})\right\rangle\\
&\ge \frac{\rho}{2N}{\rm dist}(\bar{\pmb x}^k,\mathcal{P}) + \frac{\eta}{2N} \|\nabla f(\bar{\pmb x}^k)\|^2
\end{aligned}
\end{equation}
where the last line follows from \eqref{RegCond:Real}. Combining \eqref{dist:progress}
and \eqref{InnerProd:Expe} yields
\begin{equation}\label{dist:prog:Expe}
\begin{aligned}
&\mathbb{E}\left[{\rm dist}^2(\bar{\pmb x}^{k+1},\mathcal{P})\right]\\
&\le \left(1-\frac{\rho \gamma_k}{N}\right){\rm dist}^2(\bar{\pmb x}^k,\mathcal{P})
+ \frac{\gamma_k}{2N} (\gamma_k - 2\eta)\|\nabla f(\bar{\pmb x}^k)\|^2\\
&\le \left(1-\frac{\rho \gamma_k}{N}\right){\rm dist}^2(\bar{\pmb x}^k,\mathcal{P})
\end{aligned}
\end{equation}
where the last inequality follows from $0<\gamma_k\le2\eta$. Successively
applying \eqref{dist:prog:Expe}, we get
\begin{equation}\label{dist:prog:Expe2}
\begin{aligned}
\mathbb{E}\left[{\rm dist}^2(\bar{\pmb x}^{k+1},\mathcal{P})\right]
&\le \prod\limits_{j=1}^{k} \left(1-\frac{\rho \gamma_j}{N}\right){\rm dist}^2(\bar{\pmb x}^0,\mathcal{P})\\
&\le \left(1-\frac{\rho \gamma_{\min}}{N}\right)^k{\rm dist}^2(\bar{\pmb x}^0,\mathcal{P})
\end{aligned}
\end{equation}
where $\gamma_{\min}=\min\limits_{1\le j\le k} \gamma_j$. \hfill $\Box$

\emph{Remark 4:} To guarantee convergence to the globally optimal solution, it
requires $|\alpha| \le 2\eta |\nabla f_{i_k}(\bar{\pmb x}^k)|$
or equivalently $0<\gamma_k\le2\eta$. If $\eta$ is known or can be estimated, we can perform the
one-dimensional search of \eqref{alphak:min} limited to a line segment. Note that \eqref{alphak:min}
is on minimizing a univariate quartic polynomial in an interval. This problem is easy to solve because
its solution belongs to the stationary points in the interval (if there indeed exists such a stationary point in
the interval) or the endpoints of the interval. However, $\eta$ is always not easy to estimate in practice.
From simulations, we find that dropping the box constraint $0<\gamma_k\le2\eta$ will not destroy the convergence.
This implies that the box constraint is automatically satisfied. We conjecture $\eta$ is large enough
such that $\gamma_k\le2\eta$ is always guaranteed when there are enough samples.
Therefore, this empirical observation ensures us to ignore the constraint $\gamma_k\le2\eta$
at each coordinate minimization.

\emph{Remark 5:} We only prove convergence to the global minimizer for RCD. For CCD and GCD,
theoretical proof of the convergence remains open and constitutes a future research.
Nonetheless, it is observed from the numerical simulations that the GCD converges
faster than the RCD, and CCD has comparable performance to RCD. Therefore, empirically, the GCD and CCD
also converge to the global minimum point with high probability if the sample size is large enough.

\section{CDA for Sparse Phase Retrieval with $\ell_1$-Regularization}\label{Sec:L1CD}

The CD algorithms discussed in Section \ref{Sec:CD} are applicable for general signals.
If the SOI is sparse, which is frequently encountered in practice, e.g., see \cite{Candes:SPM,GESPAR},
we can exploit the sparsity to enhance the recovery performance. In particular,
sparsity is helpful to reduce the sample number. If
$\pmb x$ is sparse, then the real-valued $\bar{\pmb x}$ is also sparse.
Inspired by the Lasso \cite{Lasso} and basis pursuit \cite{BP} in compressed sensing \cite{Candes:CS}, we adopt
the following $\ell_1$-regularization for sparse phase retrieval
\begin{equation}\label{L1Reg}
\min_{\bar{\pmb x}\in\mathbb{R}^{2N}} g(\bar{\pmb x})
:=\sum_{m=1}^{M}\left(\bar{\pmb x}^T\bar{\pmb A}_m\bar{\pmb x} - b_m\right)^2
+ \tau \|\bar{\pmb x}\|_1
\end{equation}
where $\|\bar{\pmb x}\|_1=\sum_i |\bar{x}_i|$ is the $\ell_1$-norm,
$\tau>0$ is the regularization factor and $g(\bar{\pmb x})=f(\bar{\pmb x})+\tau \|\bar{\pmb x}\|_1$.
Note that the objective function of \eqref{L1Reg} is non-differentiable due to
the non-smooth $\ell_1$-norm. As there is no gradient for \eqref{L1Reg}, the GCD is not implementable because it requires gradient for index selection. Therefore, we only discuss
the CCD and RCD for the $\ell_1$-regularization, and they are referred to as $\ell_1$-CCD and $\ell_1$-RCD, respectively. The steps of the CD for solving \eqref{L1Reg}
are similar to those in Algorithm \ref{Algo:CD}. The only difference is that an $\ell_1$-norm term is added to the scalar minimization
problem of \eqref{1D:min2}, which is shown as
\begin{equation}\label{L1:1D}
\min_{\alpha\in\mathbb{R}} \left\{\varphi(\alpha) + \tau \|\bar{\pmb x} + \alpha\pmb e_i\|_1\right\}.
\end{equation}
By ignoring the terms independent to $\alpha$, \eqref{L1:1D} is equivalent to
\begin{equation}\label{L1:1D2}
\min_{\alpha\in\mathbb{R}} \left\{\varphi(\alpha) + \tau |\alpha + \bar{x}_i|\right\}.
\end{equation}
Making a change of variable $\beta = \alpha + \bar{x}_i$, substituting
$\alpha=\beta-\bar{x}_i$ into \eqref{Obj:alpha}, and ignoring the constant term,
we obtain an equivalent scalar minimization problem
\begin{equation}\label{scalar:u}
\min_{\beta\in\mathbb{R}} \psi(\beta)
:= u_4 \beta^4 + u_3 \beta^3 + u_2\beta^2 + u_1\beta + \tau |\beta|
\end{equation}
where the coefficients of the quartic polynomial $\{u_j\}_{j=1}^4$ are calculated as
\begin{equation}\label{u:coef}
\begin{aligned}
    &u_4 = d_{4,i}\\
    &u_3 = d_{3,i} - 4\bar{x}_i d_{4,i}\\
    &u_2 = d_{2,i} - 3\bar{x}_i d_{3,i} + 6\bar{x}_i^2 d_{4,i}\\
    &u_1 = d_{1,i} - 2\bar{x}_i d_{2,i} + 3 \bar{x}_i^2d_{3,i} - 4 \bar{x}_i^3d_{4,i}.
\end{aligned}
\end{equation}
It is interesting that the solution of \eqref{scalar:u} reduces to
the well-known soft-thresholding operator in compressed sensing \cite{Wright:ST}
if $u_4 = u_3 = 0$, where the quartic polynomial reduces to a quadratic function.
Therefore, \eqref{scalar:u} is a generalization of the soft-thresholding operator
from quadratic to fourth-order functions. We call it fourth-order soft-thresholding
(FOST). Although $\psi(\beta)$ is non-smooth due to the absolute term, the closed-form
solution of its minimum can still be derived. We study the minimizer of $\psi(\beta)$ in two
intervals, namely, $[0,\infty)$ and $(-\infty,0)$. Define the set $\mathcal{S}^{+}$ containing
the stationary points of $\psi(\beta)$ in the interval $[0,\infty)$. That is,
$\mathcal{S}^{+}$ is the set of real positive roots of the cubic equation
\begin{equation}\label{psi:pos}
4u_4 \beta^3 + 3u_3 \beta^2 + 2u_2\beta + (u_1 + \tau)=0, ~\beta\ge 0.
\end{equation}
The $\mathcal{S}^{+}$ can be empty, or has one or three elements because
 \eqref{psi:pos} may have none, one, or three real positive roots.
Similarly, $\mathcal{S}^{-}$ is the set that contains
the stationary points of $\psi(\beta)$ in $(-\infty,0)$, i.e., real negative roots of
\begin{equation}\label{psi:neg}
4u_4 \beta^3 + 3u_3 \beta^2 + 2u_2\beta + (u_1 - \tau)=0, ~\beta< 0.
\end{equation}
Again, $\mathcal{S}^{-}$ can be empty, or has one or three entries.
The minimizer of $\psi(\beta)$ in $\beta\in[0,\infty)$ must be the
boundary, i.e., 0, or one element of $\mathcal{S}^{+}$.
The minimizer in $(-\infty,0)$ must be an element of $\mathcal{S}^{-}$.
In summary, the minimizer of \eqref{scalar:u} is limited to the set
$\{ 0\cup\mathcal{S}^{+}\cup\mathcal{S}^{-}\}$ which has at most seven elements, i.e.,
\begin{equation}\label{}
\beta^{\star} = \arg\min_{\beta} \psi(\beta),
~\beta\in\{ 0\cup\mathcal{S}^{+}\cup\mathcal{S}^{-}\}.
\end{equation}
Therefore, we only need to evaluate $\psi(\beta)$ over a set of at most
seven elements, whose computation is easy and simple.
The coordinate of the $\ell_1$-regularized CD is updated as
$\bar{x}_{i_k}^{k+1}\leftarrow\beta^{\star}$.
If $\mathcal{S}^{+}\cup\mathcal{S}^{-}=\emptyset$, then
$\bar{x}_{i_k}^{k+1}=\beta^{\star}=0$, which makes the solution sparse.
Certainly, even when $\mathcal{S}^{+}\cup\mathcal{S}^{-}\ne\emptyset$,
$\beta^{\star}$ may still be 0. This is why the $\ell_1$-regularized
formulation of \eqref{L1Reg}, which involves the FOST operator of \eqref{scalar:u}
at each iteration, yields a sparse solution. Clearly, $\tau$ controls the
sparseness of the solution. Generally speaking, a larger $\tau$ leads to a sparser result.

\section{Application to Blind Equalization}\label{Sec:BE}

We illustrate the application of phase retrieval to blind equalization, which is a fundamental problem in
digital communications. Consider a communication system
with discrete-time complex baseband signal model
\begin{equation}\label{}
r(n) = s(n)*h(n) + \nu(n)
\end{equation}
where $r(n)$ is the received signal, $s(n)$ is the transmitted
data symbol, $h(n)$ is the channel impulse response, $\nu(n)$ is
the additive white noise, and $*$ denotes convolution. The received
signal is distorted due to the inter-symbol interference
(ISI) induced by the propagation channel. Channel equalization is such a technique to
mitigate the ISI. Blind equalization aims at recovering the transmitted symbols without
knowing the channel response. Define the equalizer with $P$ coefficients $\pmb w = [w_0, \cdots,w_{P-1}]^T$
and $\pmb r_n = [r(n), \cdots, r(n-P+1)]^T$, the equalizer output is
\begin{equation}\label{}
y(n) = \sum_{i=0}^{P-1} w_i^* r(n-i) = \pmb w^H\pmb r_n.
\end{equation}
As many modulated signals in communications such as phase shift keying (PSK), frequency modulation (FM), and phase modulation (PM), are of constant modulus (CM), we apply the CM criterion \cite{CMA,Godard} to obtain the equalizer:
\begin{equation}\label{CM:Criterion}
    \min_{\pmb w}f_{\rm CM}(\pmb w):=\sum_{n}\left(|\pmb w^H\pmb r_n|^2-\kappa\right)^2
\end{equation}
where $\kappa>0$ is the dispersion constant defined as \cite{CMA}:
\begin{equation}\label{dispersion}
    \kappa=\frac{\mathbb{E}\left[|s(n)|^4\right]}{\mathbb{E}\left[|s(n)|^2\right]}.
\end{equation}
If $s(n)$ is of strictly constant modulus, e.g., for PSK signals, then $\kappa$ equals the square of modulus.
It is obvious that the problem of CM based blind equalization in \eqref{CM:Criterion} has the same form as the phase retrieval
of \eqref{Obj}. Both of them are multivariate quartic polynomials. The only difference between phase retrieval
and blind equalization is that the decision variable of the former is the unknown signal $\pmb x$
while that of the latter is the equalizer $\pmb w$. Therefore, the WF and CD methods can be applied
to solve \eqref{CM:Criterion}. By defining the composite channel-equalizer
response as $v(n)=h(n)*w(n)$, the quantified ISI, which is expressed as
\begin{equation}\label{ISI:def}
    {\rm ISI} =\frac{\sum_n |v(n)|^2 - \max_n |v(n)|^2}{\max_n |v(n)|^2}
\end{equation}
reflects the equalization quality. Smaller ISI implies better equalization. If ISI $=0$, then the
channel is perfectly equalized and the transmitted signal is exactly recovered up to a delay and a scalar.
Perfect equalization is only possible when there is no noise and the equalizer length $P$ is infinite for
finite impulse response (FIR) channel\footnote{The equalizer is the inverse system of the channel.
If the channel is of FIR, then its inverse has infinite impulse response (IIR). Hence, an equalizer
with infinite length is required for perfectly equalizing an FIR channel.}. Otherwise, only
approximate equalization can be achieved, which results
in a residual ISI.

\section{Simulation Results}\label{Sec:Simulation}

In our simulation study, all methods use the same initial value obtained from the spectral method \cite{WF}.
The sampling vectors $\{\pmb a_m\}$ satisfy a complex standard i.i.d. Gaussian distribution.

\subsection{Convergence Behavior}\label{SubSec:Converge}

\begin{figure}
\begin{center}
\includegraphics[width=9cm]{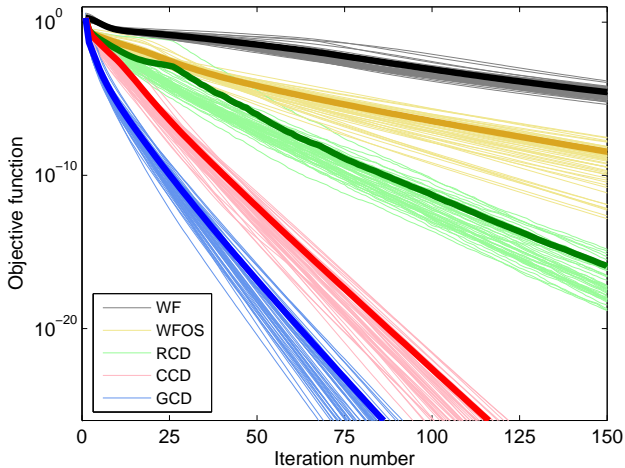}
\caption{Normalized reduction of objective function
versus number of iterations/cycles with 50 independent trials in noise-free case.}\label{Fig:obj:iter}
\end{center}
\end{figure}

\begin{figure}
\begin{center}
\includegraphics[width=9cm]{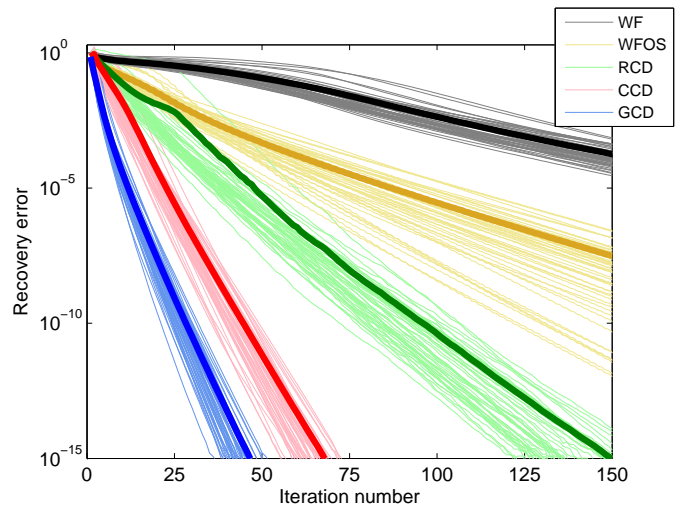}
\caption{Relative recovery error versus number of iterations/cycles
with 50 independent trials in noise-free case.}\label{Fig:NMSE:iter}
\end{center}
\end{figure}

\begin{figure}
\begin{center}
\includegraphics[width=9cm]{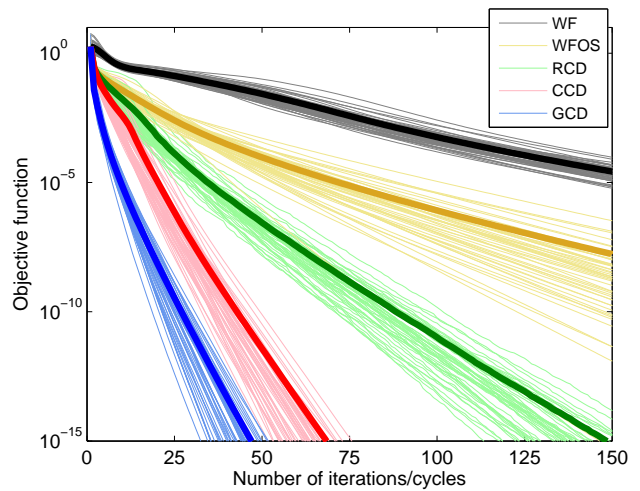}
\caption{Normalized reduction of objective function
versus number of iterations/cycles with 50 independent
trials at SNR $=$ 20~dB.}\label{Fig:obj:noisy}
\end{center}
\end{figure}

\begin{figure}
\begin{center}
\includegraphics[width=9cm]{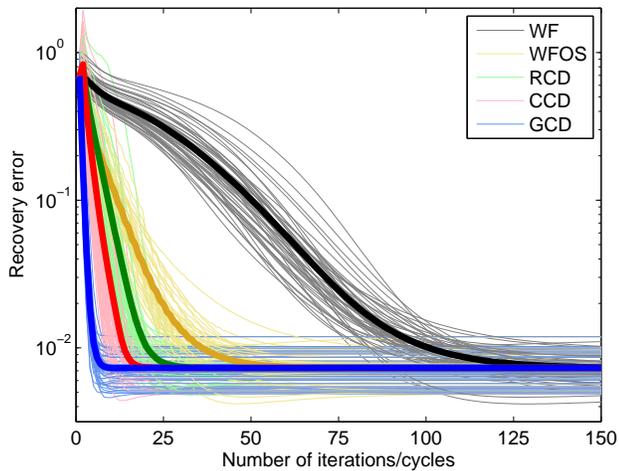}
\caption{Relative recovery error versus number of iterations/cycles
with 50 independent trials at SNR $=$ 20~dB.}\label{Fig:NMSE:noisy}
\end{center}
\end{figure}

We first investigate the convergence behavior of the three CD algorithms.
The signal $\pmb x$ and noise $\nu_m$ are i.i.d. Gaussian distributed. In this test, we set $N=64$ and $M=6N$.
The WF \cite{WF} and WFOS \cite{WFOS} that uses optimal stepsize for accelerating
the convergence speed of WF, are employed for comparison. Note that it is fair to
compare $2N$ iterations (one cycle) for the CD with one WF or WFOS iteration
because the computational complexity of the CCD and RCD per cycle is
the same as the WF per iteration. The GCD has a higher complexity for every $2N$
iterations than WF, CCD, and RCD. But still, we plot the results of GCD per cycle.
Two quantities are plotted to evaluate the convergence rate. The first quantity
is the reduction of the objective function normalized with respect to $\|\pmb b\|^2$:
\begin{equation}
\frac{f(\bar{\pmb x}^k)-f(\bar{\pmb x}^{\star})}{\|\pmb b\|^2}
\end{equation}
where $f(\bar{\pmb x}^{\star})=0$ if there is no noise. For the noisy
case, $f(\bar{\pmb x}^{\star})$ can be computed in advance to the
machine accuracy using the CD or WF method. The second quantity is
the relative recovery error, i.e.,
\begin{equation}\label{Recover:err}
\frac{{\rm dist}^2(\bar{\pmb x}^k,\mathcal{P})}{\|\bar{\pmb x}^{\star}\|^2}
\end{equation}
which reflects the convergence speed to the original signal.
Fig.~\ref{Fig:obj:iter} plots the objective reduction while Fig.~\ref{Fig:NMSE:iter} shows
the recovery error, versus the number of iterations (cycles for CD) in the absence of noise with 50 independent trials.
The averaged results are also provided with thick lines. We see that
all methods converge to the global minimum point at a linear rate. They exactly recover the true signal.
For the noisy case, the signal-to-noise ratio (SNR) in \eqref{Model} is defined as
\begin{equation}
{\rm SNR} = \frac{\mathbb{E}\left[\|\pmb b\|^2\right]}{M\sigma_{\nu}^2}
\end{equation}
where $\sigma_{\nu}^2$ is variance of $\nu_m$. Figs. \ref{Fig:obj:noisy}
and \ref{Fig:NMSE:noisy} show the normalized objective reduction and recovery
error, respectively, at SNR $=$ 20~dB. We clearly see that the three CD algorithms
converge faster than the WF and WFOS schemes. Among them, the convergence speed of the GCD is the fastest.

\subsection{Statistical Performance}

\begin{figure}
\begin{center}
\includegraphics[width=9cm]{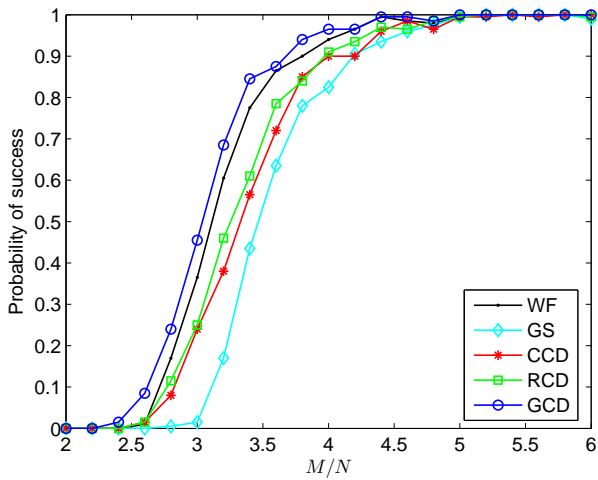}
\caption{Empirical probability of success versus number of measurements.}\label{Fig:Prob:m}
\end{center}
\end{figure}

\begin{figure}
\begin{center}
\includegraphics[width=9cm]{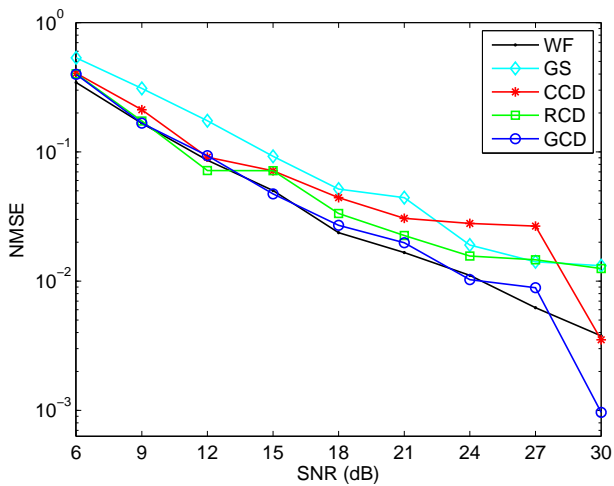}
\caption{NMSE of recovered signal versus SNR.}\label{Fig:NMSE:SNR}
\end{center}
\end{figure}

The experiment settings are the same as in Section \ref{SubSec:Converge}
except $M$ and SNR vary. The performance of the GS algorithm is also examined here.
We use the empirical probability of success and normalized mean square
error (NMSE), which is the mean of the relative recovery error in \eqref{Recover:err},
to measure the statistical performance. All results are averaged over 200 independent trials.
In the absence of noise, if the relative recovery error of a phase retrieval scheme
is smaller than $10^{-5}$, we call it success in exact recovery. Fig.~\ref{Fig:Prob:m} plots
the empirical probability of success versus number of measurements $M$. It is observed
that the GCD is slightly better than WF while CCD and RCD are slightly inferior to the WF.
Fig.~\ref{Fig:NMSE:SNR} shows the NMSE versus SNR from 6~dB to 30~dB. We see that the three CD algorithms and WF have comparable NMSEs and they are superior to the GS algorithm.

\subsection{Phase Retrieval of Sparse Signal}

\begin{figure}
\begin{center}
\includegraphics[width=9cm]{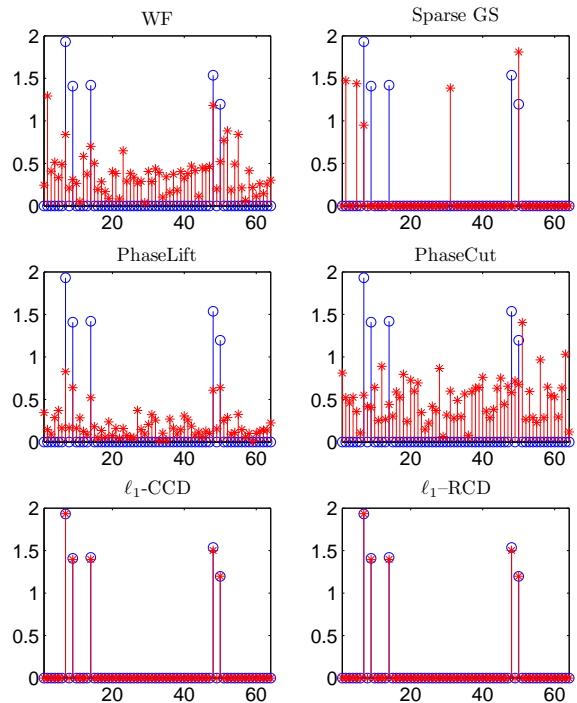}
\caption{Magnitudes of recovered signals. Red lines with star and
blue lines with circle denote the recovered and true signals, respectively.}\label{Fig:Sparse:Demo}
\end{center}
\end{figure}

\begin{figure}
\begin{center}
\includegraphics[width=9cm]{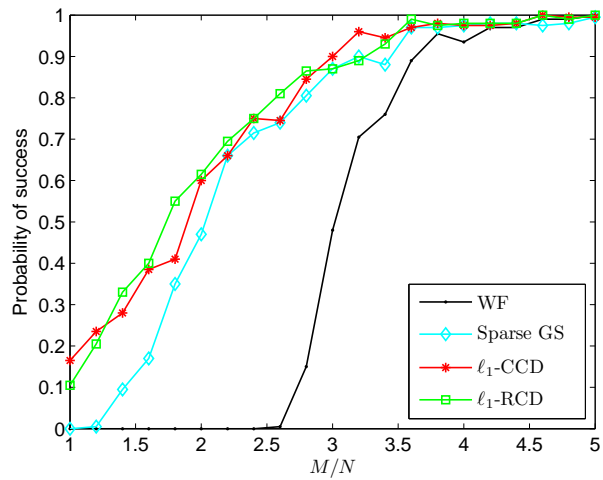}
\caption{Probability of success versus number of measurements for sparse phase retrieval.}\label{Fig:Pr:Sparse}
\end{center}
\end{figure}

In this subsection, we investigate phase retrieval of a sparse signal with $K$ nonzero elements.
In addition to WF, the two convex relaxation based methods, namely, PhaseLift \cite{PhaseLift}
and PhaseCut \cite{PhaseCut}, and sparse GS algorithm using hard-thresholding are examined for comparison.
The sparse GS algorithm needs to know $K$. We set the regularization factor as $\tau=2.35M$ for the
$\ell_1$-CCD and $\ell_1$-RCD. The support of the sparse signal is randomly selected from $[1,N]$.
The real and imaginary parts of the nonzero coefficients of $\pmb x$ are drawn as random uniform variables
in the range $\left[\frac{-2}{\sqrt{2}},\frac{-1}{\sqrt{2}}\right]\cup \left[\frac{1}{\sqrt{2}},\frac{2}{\sqrt{2}}\right]$.
Fig.~\ref{Fig:Sparse:Demo} shows the recovered signal with $K=5$ and $M=2N$ in the noise-free case. The WF, PhaseLift,
PhaseCut, and sparse GS algorithms cannot recover the signal when the sampling size $M$ is relatively small while the $\ell_1$-CCD
and $\ell_1$-RCD work well. Fig.~\ref{Fig:Pr:Sparse} plots the probability of success versus $M/N$.
By harnessing sparsity, the $\ell_1$-CCD and $\ell_1$-RCD significantly improve the recovery performance.

\subsection{Blind Equalization}

\begin{figure}
\begin{center}
\includegraphics[width=9cm]{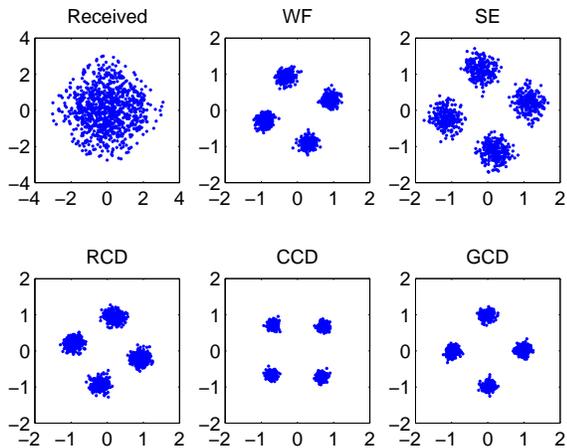}
\caption{Scatter plots of constellations of received signal and equalizer outputs.}\label{Fig:Scatter}
\end{center}
\end{figure}

\begin{figure}
\begin{center}
\includegraphics[width=9cm]{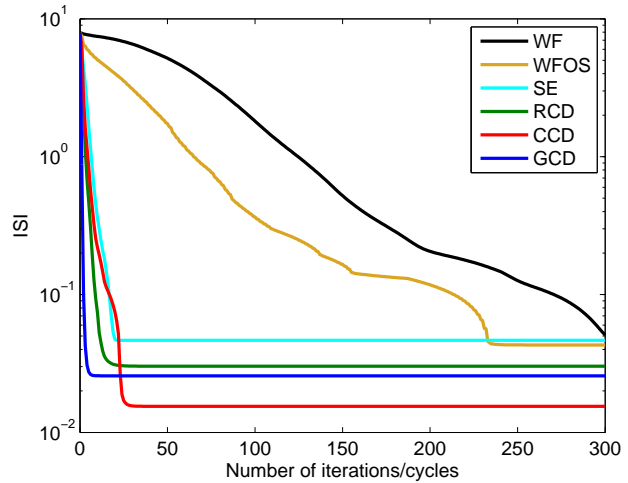}
\caption{ISI versus number of iterations/cycles.}\label{Fig:ISI:iter}
\end{center}
\end{figure}

We investigate the application of the CD and WF methods to blind equalization in the presence of white Gaussian noise.
The results of the super-exponential (SE) algorithm \cite{SEA} are also included. The transmitted
signal adopts quadrature PSK (QPSK) modulation, namely, $s(n)\in \{1,-1,{\rm j},-{\rm j}\}$.
A typical FIR communication channel with impulse response $\{0.4,1,-0.7,0.6,0.3,-0.4,0.1\}$
is adopted \cite{SEA}. Fig.~\ref{Fig:Scatter} shows the constellations of the received signal
and equalizer outputs of 1000 samples at SNR = 20~dB. We observe that the received signal
is severely distorted due to the channel propagation. The SE, WF and CD methods succeed in recovering
the transmitted signal up to a global phase rotation. We clearly see that the CCD and GCD have a higher
recovery accuracy. Fig.~\ref{Fig:ISI:iter} plots the ISI versus the number of iterations/cycles at SNR =
25~dB with 2000 samples. The ISI is averaged over 100 independent trials.
In addition to faster convergence than the WF, WFOS, and SE, the CDs (especially CCD) arrive at a lower ISI.
This means that the CDs also achieve a more accurate recovery.

\section{Conclusion}\label{Sec:Conclusion}

This paper designs CD algorithms for efficiently solving the quartic
polynomial minimization in phase retrieval. One appealing characteristic of our scheme is
the conceptual and computational simplicity: the minimum of
each one-dimensional coordinate optimization is obtained
by root finding of a univariate cubic equation, which has a
closed-form solution. Three different rules for coordinate selection
yield three CD variants, namely, CCD, RCD, and GCD. The GCD selecting
the coordinate associated with the largest absolute partial derivative
converges faster than the cyclic and randomized CDs. Theoretically, we prove
that the three CD algorithms converge to a stationary point for any initial value.
We also prove that the RCD converges to the global minimum and achieves exact
recovery with high probability provided that the sample size is large enough.
The main advantage of the CD over the full gradient methods such as WF and WFOS is its faster
convergence speed. The CD is also extended to solving the $\ell_1$-regularized quartic minimization
for phase retrieval of sparse signals. In the new application to blind equalization, the CD
can achieve lower ISI and higher recovery accuracy than several existing methods.

\section*{Acknowledgment}
W.-J. Zeng would like to thank Prof. Emmanuel J. Cand\`{e}s at Stanford University
for his suggestions and encouragement. He is also grateful to
Prof. Xiaodong Li at University of California at Davis for his explanations
on some questions of the WF method for phase retrieval.


\begin{thebibliography}{1}

\bibitem{Patent}
W.-J. Zeng and H. C. So, ``Phase retrieval using coordinate decent techniques,''
filed with U.S. Patent Office (No. 15/344,279), Nov. 2016.

\bibitem{Candes:MC}
E. J. Cand\`{e}s, Y. C. Eldar, T. Strohmer, and V. Voroninski, ``Phase retrieval
via matrix completion,''  \emph{SIAM Rev.,} vol. 57, no. 2, pp. 225--251, 2015.

\bibitem{SPM}
Y. Shechtman, Y. C. Eldar, O. Cohen, H. N. Chapman, J. Miao and M. Segev,
``Phase retrieval with application to optical imaging,''
\emph{IEEE Signal Process. Mag.,} vol. 32, no. 3, pp. 87--109, May 2015.

\bibitem{Xray}
R. P. Millane, ``Phase retrieval in crystallography and optics,''
\emph{J. Opt. Soc. Amer. A.,} vol. 7, no. 3, pp. 394--411, Mar. 1990.

\bibitem{JMiao}
J. Miao, T. Ishikawa, Q. Shen, and T. Earnest, ``Extending X-ray
crystallography to allow the imaging of noncrystalline materials, cells,
and single protein complexes,'' \emph{Annu. Rev. Phys. Chem.,} vol. 59,
pp. 387--410, 2008.

\bibitem{Nature}
B. E. Allman, P. J. McMahon, K. A. Nugent, D. Paganin, D. Jacobson,
M. Arif, and S. A. Werner, ``Imaging: Phase radiography with neutrons,''
\emph{Nature}, vol. 408, no. 6809, pp. 158--159, 2000.

\bibitem{Baykal04}
B. Baykal, ``Blind channel estimation via combining autocorrelation and blind phase estimation,'' \emph{IEEE Transactions on Circuits and Systems I}, vol. 51, no. 6, pp. 1125--1131, Jun. 2004.

\bibitem{Dainty87}
J. C. Dainty and J. R. Fienup, ``Phase retrieval and image reconstruction for astronomy,'' in \emph{Image Recovery: Theory and Application}, H. Stark, Ed., San Diego, CA: Academic Press, pp. 231--275, 1987.

\bibitem{Stefik78}
M. Stefik, ``Inferring DNA structures from segmentation data,'' \emph{Artificial Intelligence}, vol. 11, no. 1, pp. 85--114, Aug. 1978.

\bibitem{Balan}
R. Balan, P. Casazza, and D. Edidin, ``On signal reconstruction without phase,''
\emph{Appl. Comput. Harmon. Anal.,} vol. 20, no. 3, pp. 345--356, 2006.

\bibitem{STFT}
K. Jaganathan, Y. C. Eldar and B. Hassibi, ``STFT phase retrieval:
Uniqueness guarantees and recovery algorithms,'' \emph{IEEE J. Sel. Top. Signal Process.},  vol. 10, no. 4, pp. 770--781, Jun. 2016.

\bibitem{CDP}
E. J. Cand\`{e}s, X. Li, and M. Soltanolkotabi, ``Phase retrieval from coded
diffraction patterns,'' \emph{Appl. Comput. Harmon. Anal.,} vol. 39, no. 2, pp.
277--299, 2015.

\bibitem{WF}
E. J. Cand\`{e}s, X. Li, and M. Soltanolkotabi,
``Phase retrieval via Wirtinger flow: Theory and algorithms,''
\emph{IEEE Trans. Inf. Theory,} vol. 61, no. 4, pp. 1985--2007, Apr. 2015.

\bibitem{PhaseLift}
E. J. Cand\`{e}s, T. Strohmer, and V. Voroninski, ``PhaseLift: Exact and
stable signal recovery from magnitude measurements via convex programming,''
\emph{Commun. Pure Appl. Math.,} vol. 66, no. 8, pp. 1241--1274,
2013.

\bibitem{JuSun}
J. Sun, Q. Qu, and J. Wright, ``A geometrical analysis of phase retrieval,''
submitted to \emph{Found. Comput. Math.,} 2016.
Online Available: \url{http://arxiv.org/abs/1602.06664}.

\bibitem{GSA}
R. Gerchberg and W. Saxton, ``A practical algorithm for the determination
of phase from image and diffraction plane pictures,'' \emph{Optik,}
vol. 35, pp. 237--246, 1972.

\bibitem{Fienup}
J. R. Fienup, ``Phase retrieval algorithms: A comparison,''  \emph{Appl.
Opt.,} vol. 21, no. 15, pp. 2758--2769, 1982.

\bibitem{AltMin}
P. Netrapalli, P. Jain, and S. Sanghavi, ``Phase retrieval using alternating minimization,''
\emph{IEEE Trans. Signal Process.,} vol. 63, no. 18, pp. 4814--4826,
Sep. 2015.

\bibitem{Bauschke}
H. H. Bauschke, P. L. Combettes, and D. R. Luke, ``Hybrid projection--reflection
method for phase retrieval,'' \emph{J. Opt. Soc. Amer. A,} vol. 20,
no. 6, pp. 1025--1034, 2003.

\bibitem{Elser}
V. Elser, ``Phase retrieval by iterated projections,'' \emph{J. Opt. Soc. Amer. A,}
vol. 20, no. 1, pp. 40--55, 2003.

\bibitem{NP:hard}
K. G. Murty and S. N. Kabadi, ``Some NP-complete problems in quadratic and nonlinear programming,''
\emph{Math. Program.,} vol. 39, no. 2, pp. 117--129, 1987.

\bibitem{TWF}
Y. Chen and E. J. Cand\`{e}s, ``Solving random quadratic systems of equations
is nearly as easy as solving linear systems,'' \emph{Neural Information Processing Systems}, pp. 739--747, Dec. 2015, Montreal, Canada.

\bibitem{PhaseCut}
I. Waldspurger, A. d'Aspremont, and S. Mallat, ``Phase recovery, MaxCut
and complex semidefinite programming,'' \emph{Math. Program., Ser. A},
vol. 149, no. 1, pp. 47--81, Feb. 2015.

\bibitem{MaxCut}
M. X. Goemans and D. P. Williamson, ``Improved approximation algorithms
for maximum cut and satisfiability problems using semidefinite programming,''
\emph{J. ACM,} vol. 42, no. 6, pp. 1115--1145, Nov. 1995.

\bibitem{QCQP}
Z.-Q. Luo, W.-K. Ma, A. M.-C. So, Y. Ye, and S. Zhang, ``Semidefinite relaxation
of quadratic optimization problems,'' \emph{IEEE Signal Process. Mag.,} vol. 27,
no. 3, pp. 20--34, May 2010.

\bibitem{WFOS}
X. Jiang and X. Liu, ``Wirtinger flow method with optimal stepsize for phase retrieval,''
\emph{IEEE Signal Process. Lett.,}, vol. 23, no. 11, pp. 1627--1631, Nov. 2016.

\bibitem{Candes:SPM}
E. J. Cand\`{e}s and M. Wakin, ``An introduction to compressive sampling,''
\emph{IEEE Signal Proces. Mag.,} vol. 25, no. 2, pp. 21--30, Mar. 2008.

\bibitem{Candes:CS}
E. J. Cand\`{e}s and T. Tao, ``Near-optimal signal recovery from random
projections: Universal encoding strategies,'' \emph{IEEE Trans. Inf. Theory,}
vol. 52, no. 12, pp. 5406--5425, Dec. 2006.

\bibitem{SparseFienup}
S. Mukherjee and C. S. Seelamantula, ``Fienup algorithm with sparsity constraints: Application
to frequency-domain optical-coherence tomography,'' \emph{IEEE Trans. Signal Process.,}
vol. 62, no. 18, pp. 4659--4672, Sep. 2014.

\bibitem{TCai}
T. Cai, X. Li, and Z. Ma, ``Optimal rates of convergence for noisy sparse phase
retrieval via thresholded Wirtinger flow,'' \emph{Ann. Stat.}, vol. 44, no. 5, pp. 2221--2251,
Oct. 2016.

\bibitem{OMP}
J. A. Tropp and A. C. Gilbert, ``Signal recovery from random measurements
via orthogonal matching pursuit,'' \emph{IEEE Trans. Inf. Theory,} vol.
53, no. 12, pp. 4655--4666, Dec. 2007.

\bibitem{LpGreedy}
W.-J. Zeng, H. C. So, and X. Jiang, ``Outlier-robust greedy pursuit algorithms
in $\ell_p$-space for sparse approximation,'' \emph{IEEE Trans. Signal Process.,}
vol. 64, no. 1, pp. 60--75, Jan. 2016.

\bibitem{GESPAR}
Y. Shechtman, A. Beck, and Y. C. Eldar, ``GESPAR: Efficient phase retrieval
of sparse signals,'' \emph{IEEE Trans. Signal Process.,} vol. 62, no. 4, pp. 928--938,
Feb. 2014.

\bibitem{Wright}
S. J. Wright, ``Coordinate descent algorithms,''
\emph{Math. Program., Ser. A}, vol. 151, no. 1, pp. 3--34, Jun. 2015.

\bibitem{Nesterov}
Y. Nesterov, ``Efficiency of coordinate-descent methods on
huge-scale optimization problems,'' \emph{SIAM J. Optim.,}
vol. 22, no. 2, pp. 341--362, 2012.

\bibitem{Greenbaum}
A. Greenbaum, \emph{Iterative Methods for Solving Linear Systems.}
SIAM, Philadelphia, PA, 1997.

\bibitem{YeBook}
D. G. Luenberger and Y. Ye, \emph{Linear and Nonlinear Programming.}
Springer, New York, 2016.

\bibitem{Cubic:Root2}
E. T. Whittaker and G. Robinson, ``The solution of the cubic'', Section 62,
\emph{The Calculus of Observations: A Treatise on Numerical Mathematics},
4th ed. New York: Dover, pp. 124--126, 1967.

\bibitem{Beck}
A. Beck and L. Tetruashvili, ``On the convergence of block coordinate descent type methods,''
\emph{SIAM J. Optim.,} vol. 23, no. 4, pp. 2037--2060, 2013.

\bibitem{Royden}
H. L. Royden, \emph{Real Analysis.} Macmillan Publishing Company, NY, 1988.

\bibitem{Bertsekas}
D. Bertsekas, \emph{Nonlinear Programming}.
Athena Scientific, Belmont, 1999.

\bibitem{Luke}
D. R. Luke, J. V. Burke, and R. G. Lyon, ``Optical wavefront reconstruction: Theory
and numerical methods,'' \emph{SIAM Rev.,} vol. 44, no. 2, pp. 169--224, 2002.

\bibitem{Lasso}
R. Tibshirani, ``Regression shrinkage and selection via the LASSO,''
\emph{J. R. Staist. Soc. Ser. B (Methodol.),} vol. 58, no. 1, pp. 267--288, 1996.

\bibitem{BP}
S. S. Chen, D. L. Donoho, and M. A. Saunders, ``Atomic decomposition
by basis pursuit,'' \emph{SIAM Rev.,} vol. 43, no. 1, pp. 129--159, 2001.

\bibitem{Wright:ST}
S. J. Wright, R. D. Nowak, and M. Figueiredo, ``Sparse reconstruction by separable approximation,''
\emph{IEEE Trans. Signal Process.,} vol. 57, no. 7, pp. 2479--2493, Jul. 2009.

\bibitem{CMA}
C. R. Johnson, Jr., P. Schniter, T. J. Endres, J. D. Behm, D. R. Brown, and R. A. Casas, ``Blind
equalization using the constant modulus criterion: A review,''
\emph{Proc. IEEE}, vol. 86, no. 10, pp. 1927--1950, Oct. 1998.

\bibitem{Godard}
D. N. Godard, ``Self-recovering equalization and carrier tracking in two-dimensional
data communication systems,'' \emph{IEEE Trans. Commun.}, vol. 28, pp. 1867--1875, Nov. 1980.

\bibitem{SEA}
O. Shalvi and E. Weinstein, ``Super-exponential method for blind deconvolution,''
\emph{IEEE Trans. Inf. Theory,} vol. 39, no. 2, pp. 504--519, Mar. 1993.

\end{thebibliography}
\end{document}